\documentclass[a4paper,twocolumn,showpacs,amsmath,pra,amssymb]{revtex4-1} 
\usepackage[T1]{fontenc}
\usepackage[latin9]{inputenc}
\usepackage{times}
\usepackage{color}
\usepackage{xspace}
\usepackage{color,ulem}
\usepackage{amssymb,amsmath}
\usepackage{amsbsy}
\usepackage{graphicx}
\usepackage{epstopdf}
\usepackage{float}

\newcommand{\diff}{\mathrm{d}}

\newcommand{\ec}{\epsilon_{\mathrm{cut}}}

\newcommand{\etal}{\textit{et al.}}

\newcommand{\bx}{\ensuremath{\mathbf{x}}\xspace}
\newcommand{\x}{\mathbf{x}}

\renewcommand{\r}{\mathbf{r}}
\newcommand{\Cdd}{C_{\textrm{dd}}}

\newcommand{\xd}{\mathbf{x}^\prime}
\newcommand{\bk}{\mathbf{k}}

\newcommand{\ldb}{\lambda_{\rm{dB}}\xspace}
\begin{document}
\newcommand{\nt}{\tilde{n}}
\newcommand{\Lnh}{\ensuremath{\mathcal{L}}\xspace}
\newcommand{\dbx}{\diff\bx}
\newcommand{\dbk}{\diff\bk}
\newcommand{\nbe}{\ensuremath{\bar{n}_{\mathrm{BE}}}\xspace}

\title{Finite-temperature trapped dipolar Bose gas}
\author{R.~N.~Bisset} 
\author{D.~Baillie} 
\author{P.~B.~Blakie}  

\affiliation{Jack Dodd Centre for Quantum Technology, Department of Physics, University of Otago, Dunedin, New Zealand.}

\begin{abstract}
We develop a finite temperature Hartree theory for the trapped dipolar Bose gas. We use this theory  to study thermal effects on the mechanical stability of the system and density oscillating condensate states. We present results for the stability phase diagram as a function of temperature and aspect ratio. In oblate traps above the critical temperature for condensation we find that the Hartree theory predicts significant stability enhancement over the semiclassical result. Below the critical temperature we find that thermal effects are well described by accounting for the thermal depletion of the condensate.
Our results also show that density oscillating condensate states occur over a range of interaction strengths that broadens with increasing temperature.

\end{abstract}

\maketitle

\section{Introduction}  
A significant new area of interest in ultra-cold atomic gases is the study of systems in which the particles interact via a dipole-dipole interaction (DDI) \cite{Lahaye2007a}. This interest is being driven by a broad range of proposed applications from condensed matter physics to quantum information, e.g.~see \cite{Baranov2002a,*Goral2002a,*DeMille2002a,*ODell2003a,*Kawaguchi2006a,*Rabl2006a,*Buchler2007a}.
Experimental progress in the quantum degenerate regime has been driven by seminal work with $^{52}$Cr \cite{Griesmaier2005a,*Lahaye2009a,*Bismut2010a,*Pasquiou2011a}, which was Bose condensed in 2005, and more recently the realization of Bose-Einstein condensates of  $^{164}$Dy \cite{Mingwu2011a} and  $^{168}$Er \cite{Aikawa2012a}.  Polar molecules,  which have DDIs several orders of magnitude larger than those of the atomic gases, have already been produced in their ground rovibrational state \cite{Aikawa2010a,Ni2008a}, and steady progress is being made towards cooling these into the  degenerate regime. 
We also note the recent achievement of a degenerate Fermi gas of $^{161}$Dy \cite{Lu2012a}.

The DDI is long-ranged and  anisotropic with both attractive and repulsive  components.
Therefore, an important consideration is under what conditions the system is mechanically stable from collapse to a high density state.
Theoretical studies on zero temperature dipolar condensates  reveal a rich stability diagram where, 
due to the DDI anisotropy, the  stability  is strongly dependent on the geometry of the trapping potential and the properties of the short ranged (contact) interactions \cite{Yi2001a,Eberlein2005a,Ronen2007a,Lu2010a}.
 Another interesting theoretical observation is that for appropriate parameters (near instability) the condensate mode exhibits spatial oscillations and has a density maximum away from the minimum of the trapping potential \cite{Ronen2007a,Lu2010a, Wilson2009a,Wilson2011a}. However, evidence for this \textit{density oscillating} state has yet to be observed in experiment.

In this work we study the properties of a trapped dipolar Bose gas at finite temperature -- a regime largely unexplored in theory and experiments. In previous work \cite{Bisset2011} we studied the stability of a normal Bose gas (i.e.~above $T_c$) using a self-consistent semiclassical approximation. In this work we extend this study to below $T_c$ and to include quantum pressure (i.e.~beyond-semiclassical effects) by numerically solving for the condensate and  its excitations. Using this theory we study the crossover from the high temperature (above $T_c$)  to zero temperature (pure condensate) stability. Our results reveal that beyond semiclassical  effects play a significant role above $T_c$ in oblate geometry traps and enhance the stability region, and that the double instability phase diagram in this trap geometry (predicted in \cite{Bisset2011}) remains prominent. We also study the behavior of the emergent biconcave condensate (density oscillating ground state) in the finite temperature regime, and find that thermal effects enhance the density oscillation and enlarge the parameter regime over which this type of state exists. We demonstrate that the below $T_c$ temperature dependence of the stability boundary  is well-characterized by a simple model that accounts for the thermal depletion of the condensate.

\section{Formalism and numerical implementation}
\subsection{Formalism}
Here we consider a set of particles of mass $M$  confined in a cylindrically symmetric harmonic potential  \begin{equation}U_{\mathrm{tr}}(\x)=\frac{1}{2}M[\omega_{\rho}^2(x^2+y^2)+\omega_z^2z^2],\end{equation}
of aspect ratio  $\lambda=\omega_z/\omega_\rho$, with $z$ and $\rho$ representing the axial and radial directions, respectively.
We take the particles to have dipole moments polarized along the $z$ axis by an external field, such that the  DDI potential between particles is
\begin{equation}
 U_{\mathrm{dd}}(\r)=\frac{C_{\mathrm{dd}}}{4\pi}\frac{1-3\cos^2\theta}{|\r|^3},
\end{equation}
where  $C_{\mathrm{dd}}\! = \!\mu_0\mu^2~(=\!d^2\!/\epsilon_0)$  is the magnetic (electric) dipole-dipole interaction  strength and $\theta$ is the angle between the $z$ direction and the   relative separation of the dipoles ($\r=\x'-\x$). It is easy to extend our calculations to include local (contact) interactions, however here we focus on the case of pure dipole-dipole interactions, as has been realized in experiments by use of a Feshbach resonance  (e.g.~see \cite{Koch2008a}). 

The Hartree formalism we employ (see Appendix \ref{sec:appHHF} for a discussion of the relation to Hartree-Fock theory and relevant terms neglected) involves solving for the system modes using the non-local equation 
\begin{equation}
\epsilon_ju_j(\x) =\left[-\frac{\hbar^2}{2M}\nabla^2 +V_{\rm{eff}}(\x)\right]u_j(\x),\label{Hu}
\end{equation}
 where 
\begin{align}
  V_{\rm{eff}}(\x)&=U_{\mathrm{tr}}(\x)+\int  {d\xd  }\, U_{\rm{dd}}(\x-\xd)n(\xd),\label{Veff}\\
  n(\x)&=\sum_jN_j|u_j(\x)|^2, \label{nden}
 \end{align}
are the effective potential and  total density, respectively, with $N_j=[e^{\beta(\epsilon_j-\mu)}-1]^{-1}$ the equilibrium (Bose-Einstein) occupation of the mode, $\beta=1/k_BT$ the inverse temperature, and $\mu$ the chemical potential.  Equations~(\ref{Hu})-(\ref{nden}) are solved self-consistently while the chemical potential  is adjusted to ensure that the desired total number  $N=\int d^3\x\,n(\x)$ is obtained.  Below the critical temperature $T_c$ a condensate forms in the lowest mode $u_0(\x)$ with $N_0\sim N$, but the theory, as written in  Eqs.~(\ref{Hu})-(\ref{nden}), requires no additional adjustment to account for the condensate (due to our neglect of exchange) and smoothly transitions across $T_c$. We emphasize that our motivation for using this theory is that it includes the dominant direct interactions and the full discrete character of the low energy modes, yet is  more computationally efficient than Bogoliubov-based approaches. This enables us  to study  challenging problems that have not been explored, in particular   finite temperature mechanical stability, in which obtaining convergent self-consistent solutions is demanding and time consuming. Our numerical approach builds on various developments (particularly those described in \cite{Ronen2006a}) and includes a number of features to aid calculations in the finite temperature regime where interaction effects dominate (see Appendix \ref{sec:appnum} for details).
 
 The neglect of dipole exchange is consistent with other work on finite temperature bosons \cite{Ronen2007b} and zero temperature studies of fermion stability \cite{He2008a}. We would like to note that there is some justification for this approximation. Studies on a normal trapped dipolar Fermi gas  suggest that exchange interactions will quantitatively, but not qualitatively, affect stability \cite{He2008a,Zhang2010a}. Indeed, the thermodynamic study of that system presented in \cite{Baillie2010b} found that exchange interactions are typically less important than direct interactions except for traps that are close to being isotropic. Similarly, Ticknor studied the quasi-two-dimensional Bose gas using the  Hartree-Fock-Bogoliubov-Popov (HFBP) meanfield theory \cite{Ticknor2012a} and found that exchange terms were generally less important than direct terms.

\section{Results} 
\subsection{Comparison to previous calculations}

To benchmark our Hartree calculations we perform a quantitative comparison to the  HFBP  calculations that Ronen \etal~\cite{Ronen2007b} performed for the three-dimensional trapped Bose gas at finite temperature.   In  Secs.~\ref{seccondfrac} and \ref{secDenosc} we make this comparison for two different sets of results from \cite{Ronen2007b}.

We note that those HFBP calculations excluded  \textit{thermal exchange} interactions, although they did include \textit{condensate exchange} interactions (exchange interaction of condensate atoms on the thermal excitations) \footnote{See Appendix \ref{sec:appHHF} for definitions of these exchange terms.}. We extended our Hartree algorithm to include condensate exchange but found it made negligible difference to the predictions and do not include results with this term here.

\subsubsection{Condensate Fraction}\label{seccondfrac}
\begin{figure}[!tbh]
\begin{center}
\includegraphics[width=3.2in]{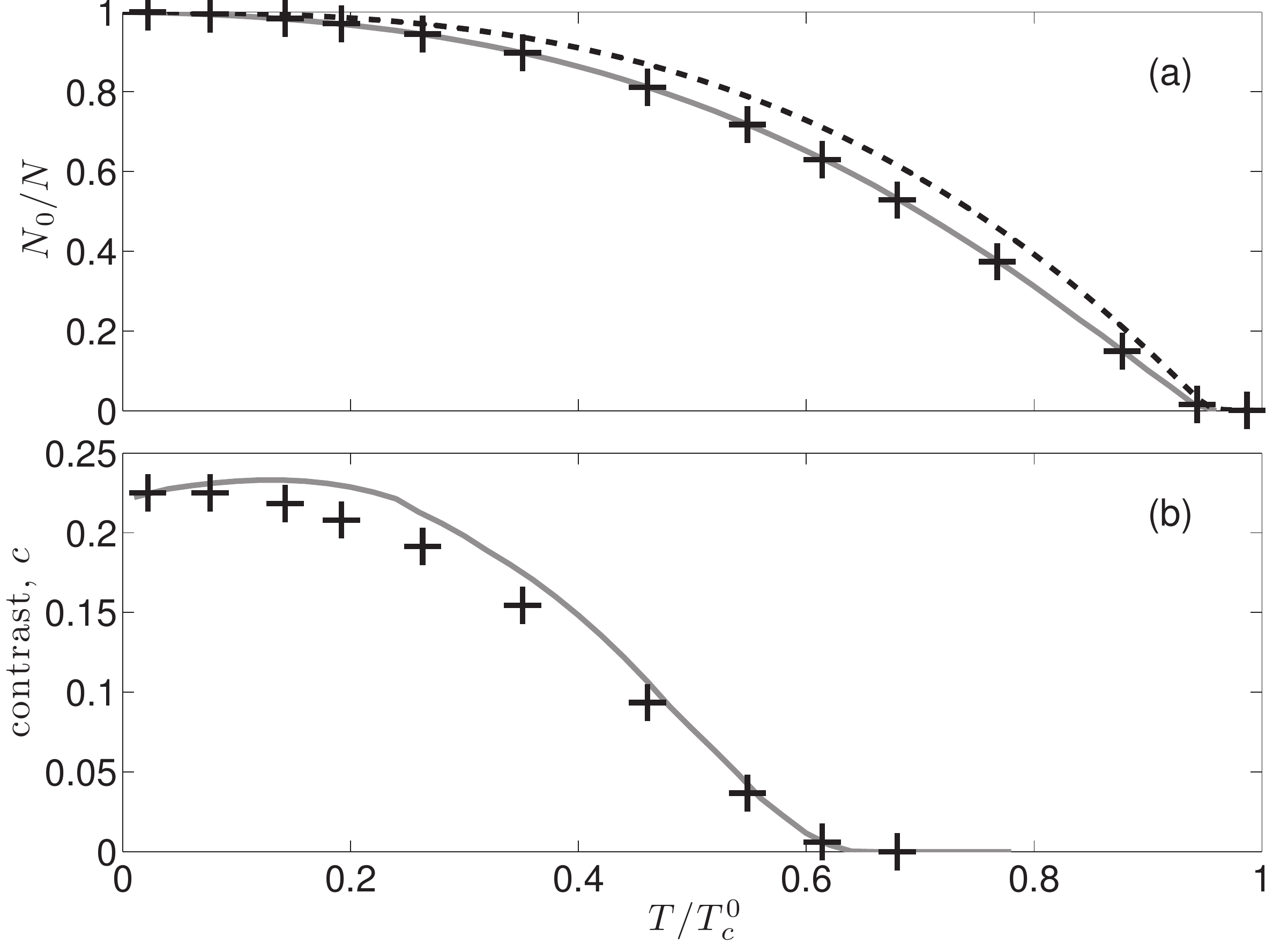} 
\caption{(a) Condensate fraction and (b) density oscillation contrast   (see text) for a dipolar BEC in a $\lambda=7$ pancake trap.  Hartree results (pluses), HFBP results (solid lines), ideal gas result (dashed line). 
HFBP data corresponds to results shown in Figs.~5 and 6 of Ref.~\cite{Ronen2007b}.
Other parameters:   $\{\omega_\rho,\omega_z\}=2\pi\times\{100,700\}\,\mathrm{s}^{-1}$, $N = 16.3\times10^3\,$  $^{52}$Cr atoms with contact interactions tuned to zero. $T_c^0 =\sqrt[3]{N/\zeta(3)}\hbar\omega/k_B$ is the ideal condensation temperature, where $\omega=\sqrt[3]{\omega_{\rho}^2\omega_z}$ and $\zeta(\alpha)$ is the Riemann zeta function with $\zeta(3)\approx1.202$.
 \label{Fig:HFBCompare}}
\end{center}\vspace{-0.4cm}
\end{figure}

The results of the first comparison we perform are presented in Fig.~\ref{Fig:HFBCompare}(a). There we compare the condensate fraction, as a function of temperature, for a system with $\lambda=7$. We observe that the Hartree and HFBP theories predict an appreciably lower condensate fraction than the ideal case, and are in very good agreement with each other over the full temperature range considered. The low energy excitations of a Bose-Einstein condensate are quasi-particles, which are accurately described by Bogoliubov theory (such as the HFBP theory), however the thermodynamic properties of the system are dominated by the single particle modes (e.g.~see \cite{Dalfovo1997a}). A  comparison of the Bogoliubov and Hartree-Fock spectra of a $T=0$ dipolar Bose-Einstein condensate (BEC)  was made in \cite{Ronen2006a}. That comparison revealed that the spectra were almost identical, except for low energy modes with low values of angular momentum, where small differences in the mode frequencies were observed.  

\subsubsection{Density Oscillating Ground States}\label{secDenosc}

An interesting feature of dipolar condensates is the occurrence of ground states with density oscillation features, where the condensate density has a local minimum at trap center.  For a cylindrically symmetric trap these states are biconcave (red blood cell shaped -- surfaces of constant density are shown in Fig.~\ref{Fig:Biconcave}) first predicted for $T=0$ condensates in Ref.~\cite{Ronen2007a}. In the purely dipolar case such biconcave states occur under certain conditions of trap and dipole parameters, but notably only for $\lambda\gtrsim6$ and for dipole strengths close to instability. In \cite{Ronen2007b} the HFBP technique was used to assess the effect of temperature on the density oscillating states.  This was characterized by the contrast, a measure of the magnitude of the density oscillation, defined as
\begin{equation}
c=1-\frac{n(\mathbf{0})}{n_{\rm{max}}},\label{contrast}
\end{equation}
where $n(\mathbf{0})$ is the density at trap center and $n_{\rm{max}}$ is the maximum density of the system.

In Fig.~\ref{Fig:HFBCompare}(b) we compare our Hartree and HFBP theories for the contrast. 
This comparison reveals some small residual differences between the theories, however the results are in reasonable agreement and both predict that the contrast goes to zero (i.e.~the condensate returns to having maximum density at trap center) at  $T\approx0.65T_c^0$.

\subsection{Mechanical stability}
Our first application of the Hartree theory is to study the finite temperature mechanical stability of a trapped dipolar Bose gas. To do this we construct a phase diagram for the range of dipole strengths for which the gas is stable for a number of different trap geometries.
 Such stability  properties, and the dependence on interactions and trap geometry, have been measured accurately in the dipolar system in the zero temperature limit (e.g.~see \cite{Koch2008a}). We note theoretical studies \cite{Houbiers1996,Bergeman1997,Davis1999} showing the important role of temperature on the observed 
  stability of  $^{7}$Li condensates \cite{Bradley1995a}, which have an attractive contact interaction.

\subsubsection{Locating the stability boundary}\label{sec:locateinstability}
\begin{figure}[!tbh]
\begin{center} 
\includegraphics[width=3.5in]{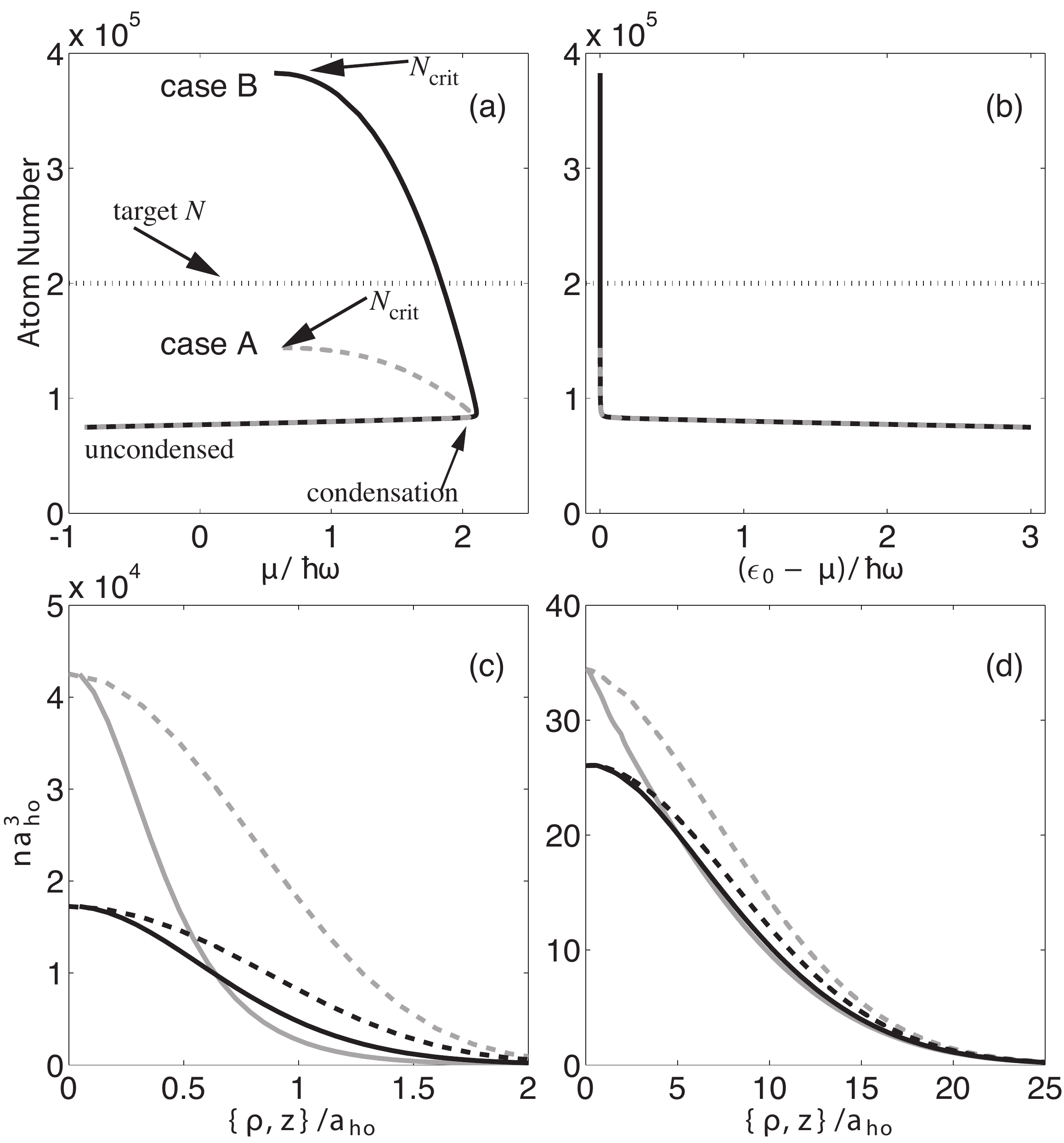}
\caption{Locating instability (upper subplots):
(a) The total number of atoms of the self-consistent  Hartree solution  versus chemical potential for $\lambda = 1/8$, $k_BT= 40\hbar\omega $ and $C_{\rm{dd}} = 7\times10^{-4}\hbar\omega a_{\mathrm{ho}}^3~\rm{(dashed, case~A)},1.5\times10^{-4}\hbar\omega a_{\mathrm{ho}}^3~\rm{(solid, case~B)}$, with $a_{\rm{ho}}=\sqrt{\hbar/M\omega}$. Each line terminates at the point of instability and occurs at the respective critical number $N_{\mathrm{crit}}$. (b) Same results as in (a) but plotted against $\epsilon_0-\mu$. The dotted line represents the target number, in this case $N=2\times 10^5$. 
Density Profiles (lower subplots): Solid (dashed) line represents the radial (axial) density $n$, higher curves are near the stability boundary. $\lambda = 1$, $N=2\times10^5$.
(c) $T/T_c^0=0.82$ ($N_0/N\approx0.43$) and $\Cdd/4\pi C_0 = \{3.65\times10^{-4}$ (gray), $1.83\times10^{-4}$ (black)$\}$. (d) $T/T_c^0=1.27$ and $\Cdd/4\pi C_0 = \{2.91$ (gray), $  1.22$ (black)$\}$. 
We have introduced the interaction strength unit
$C_0= {\hbar\omega a_{\rm{ho}}^3}/\sqrt[6]{N}$ which is convenient for cases where $N$ is fixed, and allows our subsequent results to be directly compared to those in \cite{Bisset2011}. 
 \label{Fig:dpimuN}}
\end{center}\vspace{-0.4cm}
\end{figure}
 
We consider a trapped sample of fixed mean number $N$ and wish to determine the values of the dipole interaction parameter for which the system is mechanically stable as a function of temperature.  In doing so we construct a phase diagram in $\{\Cdd,T\}$-space that indicates the stable region. In practice we locate the stability boundary (i.e.~a curve) that separates the stable and unstable regions. Our procedure to obtain this boundary involves a computationally intensive search through parameter space to find the self-consistent solutions on the verge of instability.  
Determining the stability boundary for fixed mean number $N$ complicates this process: since we work in the grand-canonical ensemble where the proper variables are $\{\mu, \Cdd,T\}$, an additional iterative search over the parameter $\mu$ is required to fix $N$ to the desired target number. 

In Fig.~\ref{Fig:dpimuN} we provide some examples to illustrate how we identify the value of the DDI at the stability boundary  for a gas with (target number) $N=2\times10^5$ atoms at a particular temperature. 
To do this we show the dependence of total atom number on $\mu$  for two different values of the DDI [Fig.~\ref{Fig:dpimuN}(a)]. For both curves the total number increases as we move along these curves until some maximum value $N_{\mathrm{crit}}$ is reached at which the system becomes unstable.  The non-monotonic behavior of these curves arises because the ground state energy $\epsilon_0$ changes as the number of atoms increases, and hence the role of DDIs increases. For this reason we also show the same two cases, but as a function of $\epsilon_0-\mu$, in Fig.~\ref{Fig:dpimuN}(b).

The sharp cusps in  Figs.~\ref{Fig:dpimuN}(a) and (b) correspond to the point where the system condenses [i.e.~where $\epsilon_0-\mu\approx0$]. 
The dependence of $\epsilon_0$ on $N_0$ is strongly dependent on the trap geometry, and for the cases  we consider here with $\lambda=1/8$, $\epsilon_0$ decreases with increasing $N_0$. This is because the head-to-tail character, in the cigar geometry, emphasizes the attractive part of the DDI so that as the condensate number increases, $\epsilon_0$ ($\approx\mu$) decreases. 

For case A  in Fig.~\ref{Fig:dpimuN}(a)  the number at which collapse occurs is less than the target number, thus we conclude that the DDI used in this calculation lies within the unstable region for the system (i.e.~no stable solution can be found for $N$ atoms with this value of DDI). In contrast, for case B  in Fig.~\ref{Fig:dpimuN}(a) $N_{\mathrm{crit}}>N$ and thus the value of DDI is in the  stable region. To locate the stability boundary points we need to trace out these curves for various  ${\Cdd}$ values until we find  $N_{\mathrm{crit}}=N$ to within our numerical tolerance (this has to be done for each value of $T$).  This process is painstaking and can take several days to find a single point on the stability boundary. 

We identify the self-consistent Hartree solutions as being unstable when they become grid-dependent. This means that as the distance between grid points tends to zero, the radial width of the cloud contracts and the chemical potential tends to negative infinity. Precisely locating the instability point is a stringent numerical task and requires careful convergence tests. For condensates with contact interactions this type of \textit{numerical instability} analysis was applied in Refs.~\cite{Davis1999,Bergeman1997,Houbiers1996} (also see Ref.~\cite{Bisset2011}). 
In Fig.~\ref{Fig:dpimuN}(a) the instability point occurs at the end of the upper horizontal plateau in the $N$ versus $\mu$ curves (compare to Fig.~1 of \cite{Houbiers1996}). 
We show examples of the spatial density profiles for a spherical trap in Figs.~\ref{Fig:dpimuN}(c) and (d). The system considered in Fig.~\ref{Fig:dpimuN}(c) is condensed, while that considered in Fig.~\ref{Fig:dpimuN}(d) is above the critical temperature. For both cases a result is shown that is well inside the stable region (black curves) and near the stability boundary (gray curves). Despite a large difference in the density scales of the two regimes they both exhibit a similar sharpening of the density profile  near instability.

An additional consideration emerges for stability calculations below $T_c$ in regimes where the condensate is in a density oscillating state. Here the first mode to go soft (and then develop imaginary parts) as the stability boundary is reached is a $m\ne0$ quasi-particle mode \cite{Wilson2009}, where $m$ is the angular momentum projection quantum number (so called angular roton mode \cite{Ronen2007a}). This instability is not revealed in the Hartree excitations, and as we solve for the condensate in the $m=0$ space (see Appendix \ref{sec:appnum}), the condensate does not exhibit numerical instability.  Thus in cases where the condensate exhibits a density oscillating state we  perform a Bogoliubov analysis of the condensate mode (within the effective potential of the self-consistent Hartree solution) to determine if any $m\ne0$ modes have become unstable \footnote{For an analogous calculation, but at $T = 0$, see \cite{Ronen2006a}}.

\subsubsection{Stability above $T_c$}\label{sec:stababoveTc}

\begin{figure}[!tbh]
\begin{center}
    \includegraphics[width=3.2in]{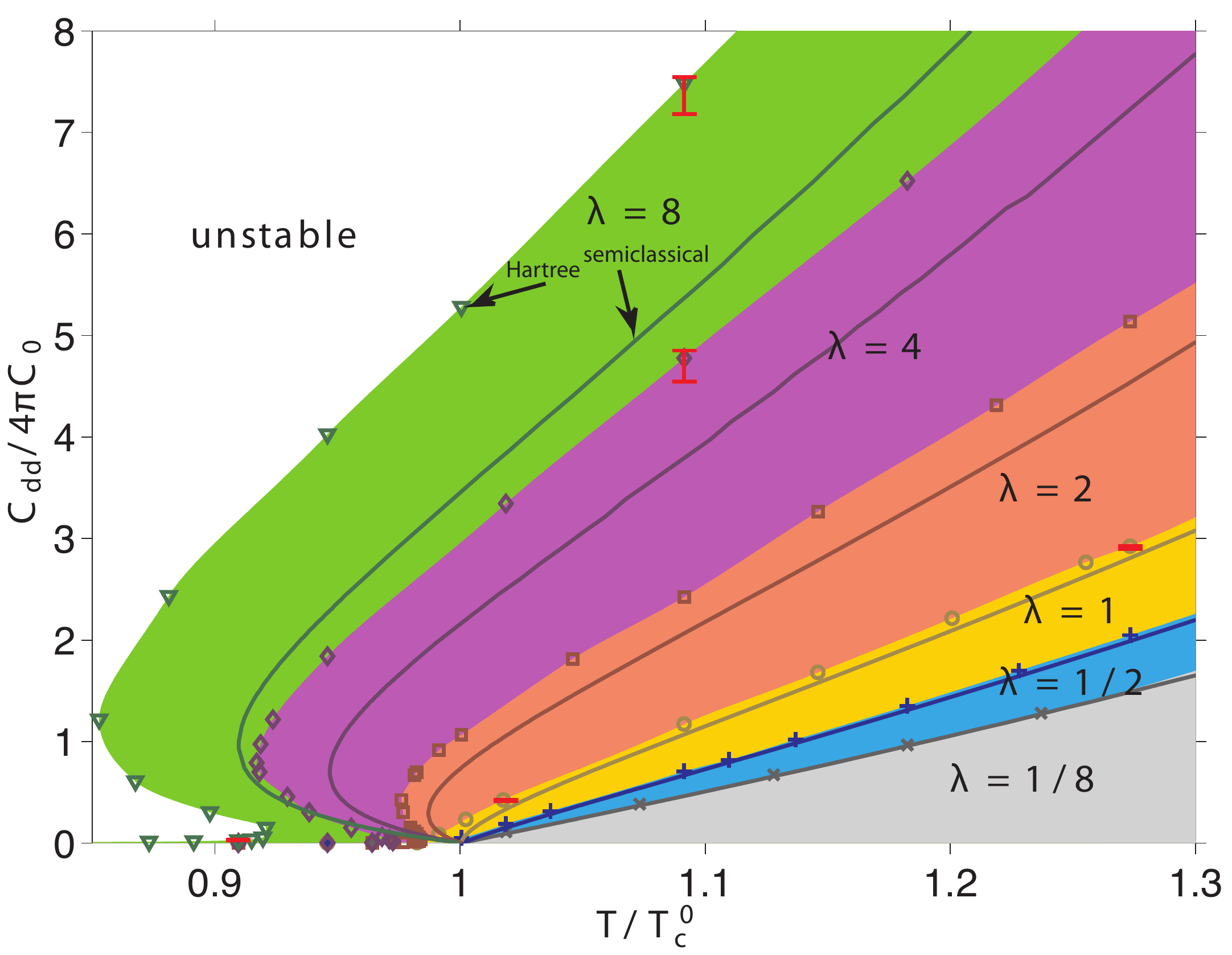}
\caption{(Color online) Stability regions in DDI-temperature space. Shaded regions indicate stability for each geometry, from top to bottom $\lambda = $ \{8, 4, 2, 1, 1/2, 1/8\}, the geometric mean trap frequency is fixed and $N=2\times10^5$. Actual data points represented by symbols while the shading of the stable regions interpolates to guide the eye, the semiclassical model is given by the solid curves.  Error bars represent the 1 $\sigma$ spread in the convergence test (see Appendix \ref{Sec:ErrorBars} for more details). 
 \label{Fig:Stab}}
\end{center}\vspace{-0.4cm}
\end{figure}

In   Fig.~\ref{Fig:Stab} we show our results for  the stability of the normal phase.  In previous work we examined this regime using a semiclassical Hartree approach in which the density is 
\begin{equation}
n(\x)=\ldb^{-3}\zeta_{3/2}\left(e^{\beta[\mu-V_{\rm{eff}}(\x)]}\right),\label{den}
\end{equation}
where  $V_{\rm{eff}}(\x)$ is the effective potential calculated using $n(\x)$ [see Eq.~(\ref{Veff})], $\zeta_{\alpha}(z)=\sum_{j=1}^{\infty} z^j/j^\alpha$ is the Bose  function, and  $\ldb=h/\sqrt{2\pi Mk_BT}$. 
The semiclassical results are shown as solid lines in Fig.~\ref{Fig:Stab}. 

We observe that as a general trend the stability region grows with increasing $\lambda$. The strong geometry dependence of these results arises from the anisotropy of the dipole interaction: In oblate geometries ($\lambda>1$) the dipoles are predominantly side-by-side and interact repulsively (stabilizing), whereas in prolate geometries ($\lambda<1$) the attractive (destabilizing) head-to-tail interaction of the dipoles dominates (a similar geometry dependence is observed for the stability of  $T=0$  dipolar condensates \cite{Ronen2007a,Lu2010a}).

A primary concern is the nature of \textit{beyond semiclassical} effects, i.e.~what differences emerge from our diagonalized Hartree theory over the semiclassical formulation.  Most prominently in the results of Fig.~\ref{Fig:Stab} we observe that while the Hartree and semiclassical stability boundaries are in good agreement for prolate geometries, in oblate traps the Hartree results are significantly more stable. This difference between the boundaries predicted by the two theories increases with increasing $\lambda$. 
This observation is   surprising because our calculation is for a rather large number of atoms ($N=2\times10^5$), where the semiclassical approximation would normally  be expected to furnish an accurate description of the above $T_c$ behavior.

We attribute this failure of the semiclassical theory to its inappropriate treatment of the interactions between the low energy modes \footnote{For definiteness, this discussion relates to the in-plane   interaction  between atoms in the lowest $z$ vibrational mode.}. The nature of the DDI, when tightly confined  along the polarization direction, has been extensively studied in application to pure BECs \cite{Fischer2006a,Wilson2011a}, where it has been shown  that it confers additional stability on the system, as verified in recent experiments \cite{Muller2011a}. 
 This  arises from a confinement induced momentum dependence of the interaction: the interaction is repulsive (stabilizing) for low momentum interactions, but decays to being attractive with a characteristic wavevector  $k\sim 1/a_z$ set by the $z$ confinement length   $a_z=\sqrt{\hbar/M\omega_z}$. 
 Notably these features of the confined interaction mediate  BEC instability through the softening of radially excited modes with a wavelength $\sim a_z$ \cite{Santos2003a,Fischer2006a,Wilson2010a,Ticknor2011a,Blakie2012a}.

It is not clear that these confinement effects will be applicable at a modestly oblate trap with $\lambda=8$, however numerical studies have revealed that quasi-particle modes with a wavelength $\sim a_z$ soften in a BEC with $\lambda=7$ \cite{Ronen2007a}.  
Within the limited range of results we have for $\lambda>1$ we see evidence consistent with confinement induced effects playing an important role in the above $T_c$ Hartree calculations. Notably, that  the relative difference between the stability boundaries of the Hartree and semiclassical calculations scale with $1/a_z^2$.  Also, when the system is unstable, during the self-consistency iterations (prior to collapse) strong radial density fluctuations develop in the system 

 A key prediction from our semiclassical study \cite{Bisset2011} is a double instability feature in oblate trapping geometries arising from the interplay of thermal gas saturation and the anisotropy of the DDI. Our Hartree calculations in this oblate regime, despite shifting the stability boundary from the semiclassical prediction by a considerable amount, reveal that the double instability feature  is robust to beyond-semiclassical effects.

A prominent feature of the semiclassical calculation  is that the stability curves for the purely dipolar gas terminate at the critical point with $\Cdd=0$ (i.e.~predicting that without contact interactions only an ideal gas is stable below $T_c$). This occurs because the local compressibility at trap center diverges at the critical point and the gas is unstable to any attractive interaction (see \cite{Bisset2011}). In the  beyond-semiclassical  calculations the trap provides a finite momentum cutoff that prevents the divergence of compressibility, and thus the system has a finite residual stability at and below $T_c$ (which we consider in Sec.~\ref{sec:stablebelowTc}).

\subsubsection{Stability below $T_c$}\label{sec:stablebelowTc}

\begin{figure}[!tbh]
\begin{center}
    \includegraphics[width=3.2in]{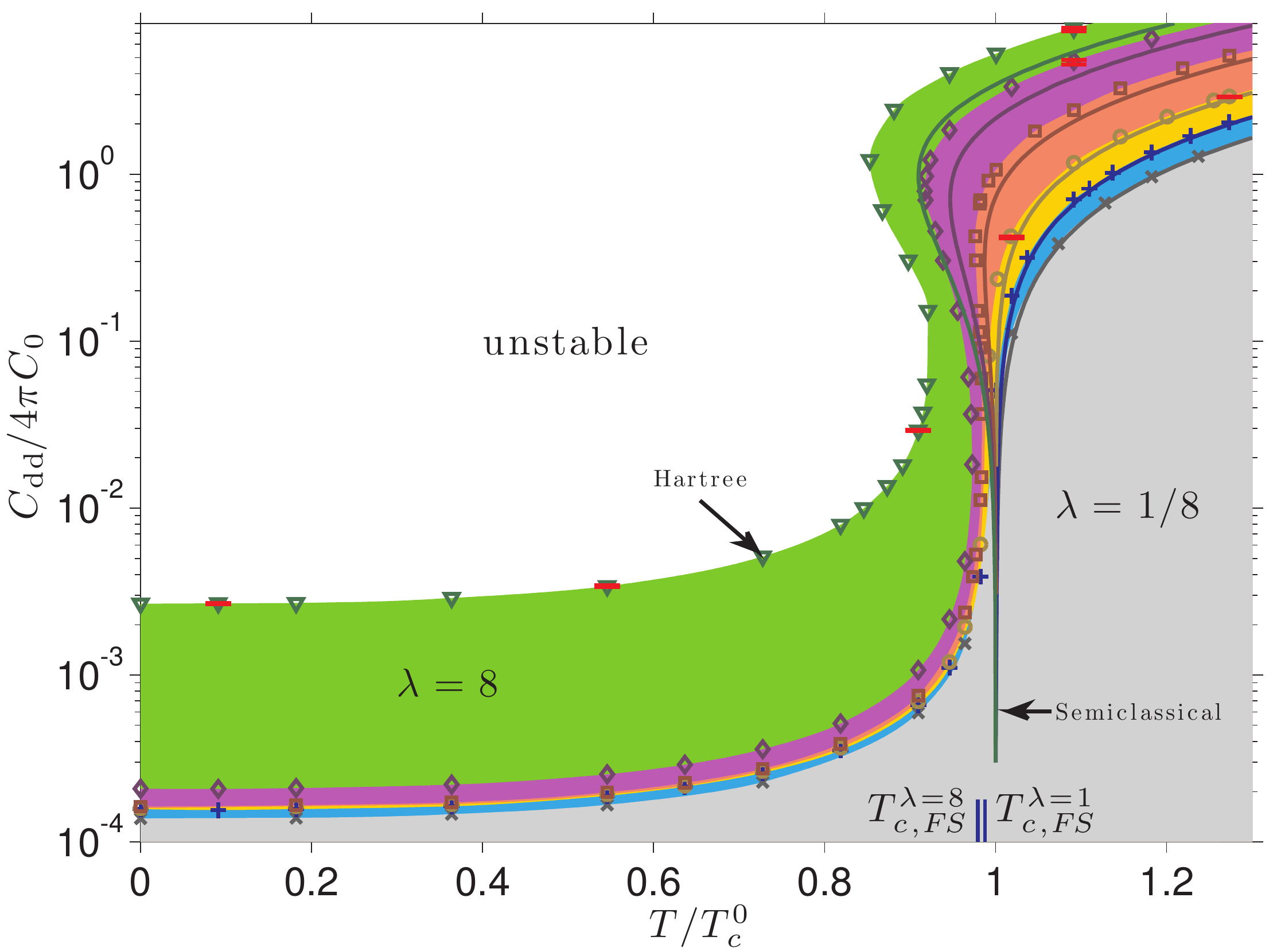}
\caption{(Color online) Stability boundary focusing on  the below $T_c$ behavior (line styles as in Fig.~\ref{Fig:Stab}). For reference the ideal finite size adjusted critical temperature $T_{c,FS}$  \cite{Dalfovo1999} for two geometries ($T_{c,FS}^{\lambda=1}$ and $T_{c,FS}^{\lambda=8}$) are indicated by short vertical lines.  The effect of DDIs on $T_c$ was calculated perturbatively in \cite{Glaum2007,Glaum2007A}, however our results are far outside the perturbative regime.
 \label{Fig:StabLog}}
\end{center}\vspace{-0.4cm}
\end{figure}

In Fig.~\ref{Fig:StabLog} we consider the stability below $T_c$ where the semiclassical model does not apply. 
 These results are identical to those shown in Fig.~\ref{Fig:Stab}, but the below $T_c$ details are revealed using a logarithmic vertical axis. Compared to the above $T_c$ gas the condensate is rather fragile, with the critical DDI strength   defining the stability boundary decreasing by $\sim$3 to 4 orders of magnitude.

 \begin{figure}[!tbh]
\begin{center}
    \includegraphics[width=3.2in]{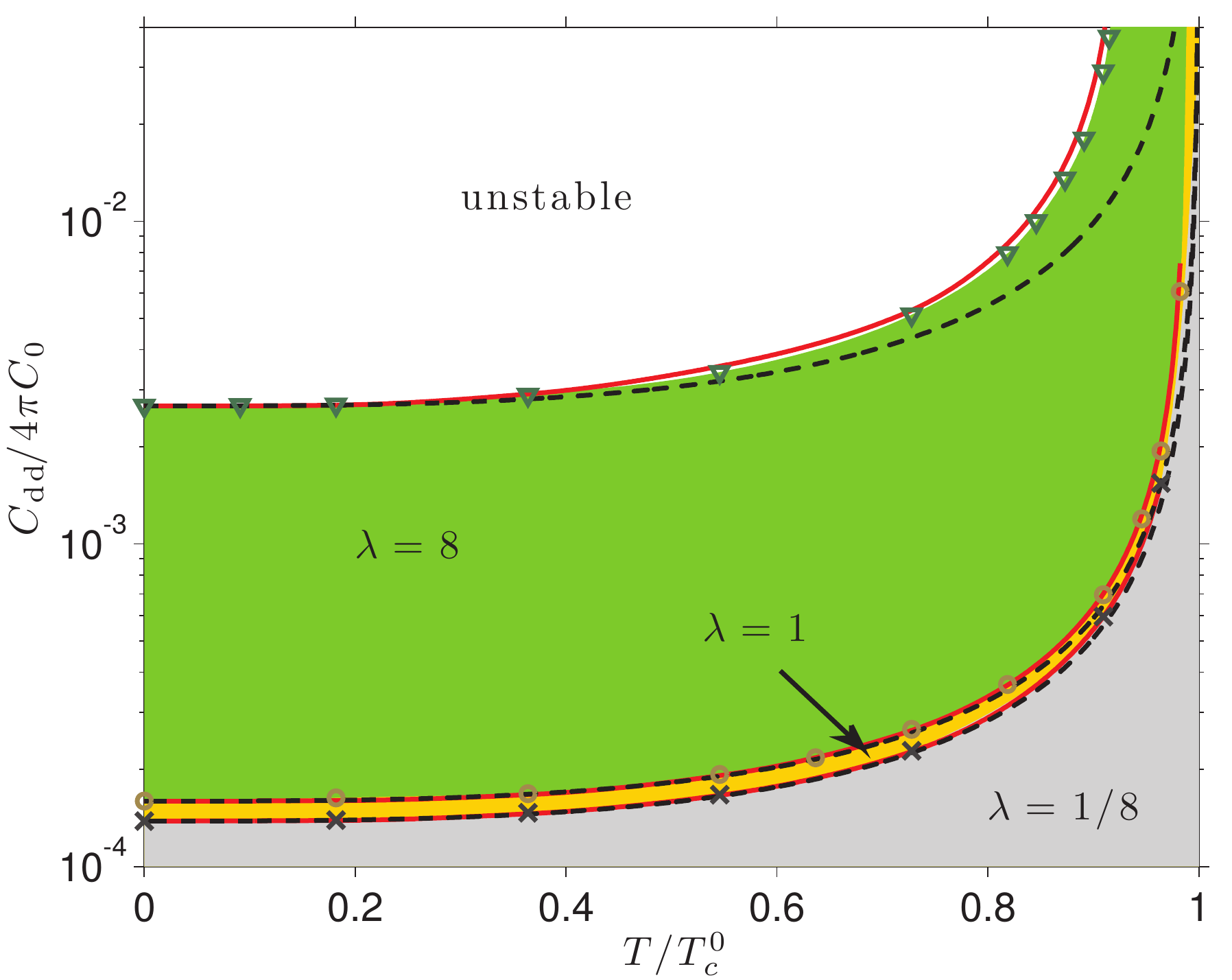}
\caption{(Color online) Stability boundary scaling. The stability boundary results (symbols) have been taken from Fig.~\ref{Fig:StabLog} for $\lambda$ = \{8,1,1/8\} (top to bottom). Dashed line prediction is based on a non-interacting $N_0$ scaling (see text) and the solid line uses the $N_0$ calculated from the Hartree solutions.
 \label{Fig:StabScalingLog}}
\end{center}\vspace{-0.4cm}
\end{figure}

In the zero temperature limit our results agree with previous calculations based on solving the Gross-Pitaevskii equation \cite{Ronen2007a}. 
This agreement is expected as the two theories are identical when  the excited modes have vanishing population.
For a pure condensate, the  critical DDI strength depends on the condensate number and trap geometry according to \cite{Ronen2007a}
\begin{equation}
\Cdd^{\star}=\frac{F(\lambda)}{N_0},\qquad{(T=0)}\label{Dcrit}
\end{equation}
with $F(\lambda)$ a rather interesting function of trap geometry alone, as   characterized in Fig.~1 of \cite{Ronen2007a} \footnote{Note  we use fixed geometric mean trap frequency, whereas  \cite{Ronen2007a} fixes $\omega_{\rho}$. The interaction parameter used in \cite{Ronen2007a} ($N\gg1$ limit) is $D= N\Cdd/4\pi\hbar\omega_\rho a_\rho^3$) with $a_{\rho}=\sqrt{\hbar/M\omega_{\rho}}$, which relates to our parameter as $D=N^{5/6}\lambda^{-1/6}\Cdd/4\pi C_0 $.}. {More generally, beyond the case of pure DDIs,  $F$  also depends on the contact interaction strength,~e.g.~see \cite{Lu2010a,Blakie2012a}}.

As temperature increases, but focusing on $T<T_c^0$, we observe in Fig.~\ref{Fig:StabLog} that the stability boundary  increases significantly. 
This occurs  because as the temperature increases the condensate is thermally depleted. Indeed,  by simply  accounting for the thermal depletion we can immediately  extend result (\ref{Dcrit}) to predict the critical value of the DDI  at finite temperature $\Cdd^{\star}(T)$:
\begin{equation}
\Cdd^{\star}(T)=\frac{F(\lambda)}{N_0(T)}=\Cdd^{\star}(0) \frac{N}{N_0(T)},\label{simplemodel}
\end{equation}
where the last expression is obtained using   $N_0(T=0)=N$.  
Equation (\ref{simplemodel}) predicts that the stability at finite temperature increases inversely proportional to the condensate occupation and, as shown in ~Fig.~\ref{Fig:StabScalingLog}, provides a good description of the stability predictions from the full Hartree calculations. In these comparisons we have used two models for the condensate occupation: (i) the non-interacting  prediction  
\begin{equation}
N^{\mathrm{NI}}_0(T)=N[1-(T/T_c^0)^3],\label{N0ni}
\end{equation}
and (ii) the value of $N_0(T)$ obtained from the Hartree calculations. Equation (\ref{simplemodel}) using $N^{\mathrm{NI}}_0(T)$ provides a good prediction for Hartree stability curves with $\lambda= 1/8$ and $1$. For the oblate system ($\lambda=8$) agreement is not as good as is apparent in Fig.~\ref{Fig:StabLog} for $T\gtrsim0.5T_c^0$. In this case the values of $\Cdd$ at the stability boundary are much higher than for the other values of $\lambda$, and thus interaction effects more significantly affect the condensate. However,  much better agreement is obtained if we take $N_0(T)$ from the Hartree solution. 

We note that the simple model (\ref{simplemodel}) does not account for any other effects of the thermal cloud on the condensate [e.g.~thermal back action through modifications of $V_{\mathrm{eff}}(\x)$].
 Thus, the level agreement of this simple model with the full Hartree results suggest that these additional effects are not significant in the regimes we have studied.

\subsection{Thermal effects on biconcavity}\label{Sec:ThermBicon}
 
\begin{figure}[!tbh]
\begin{center} 
    \includegraphics[width=3.5in]{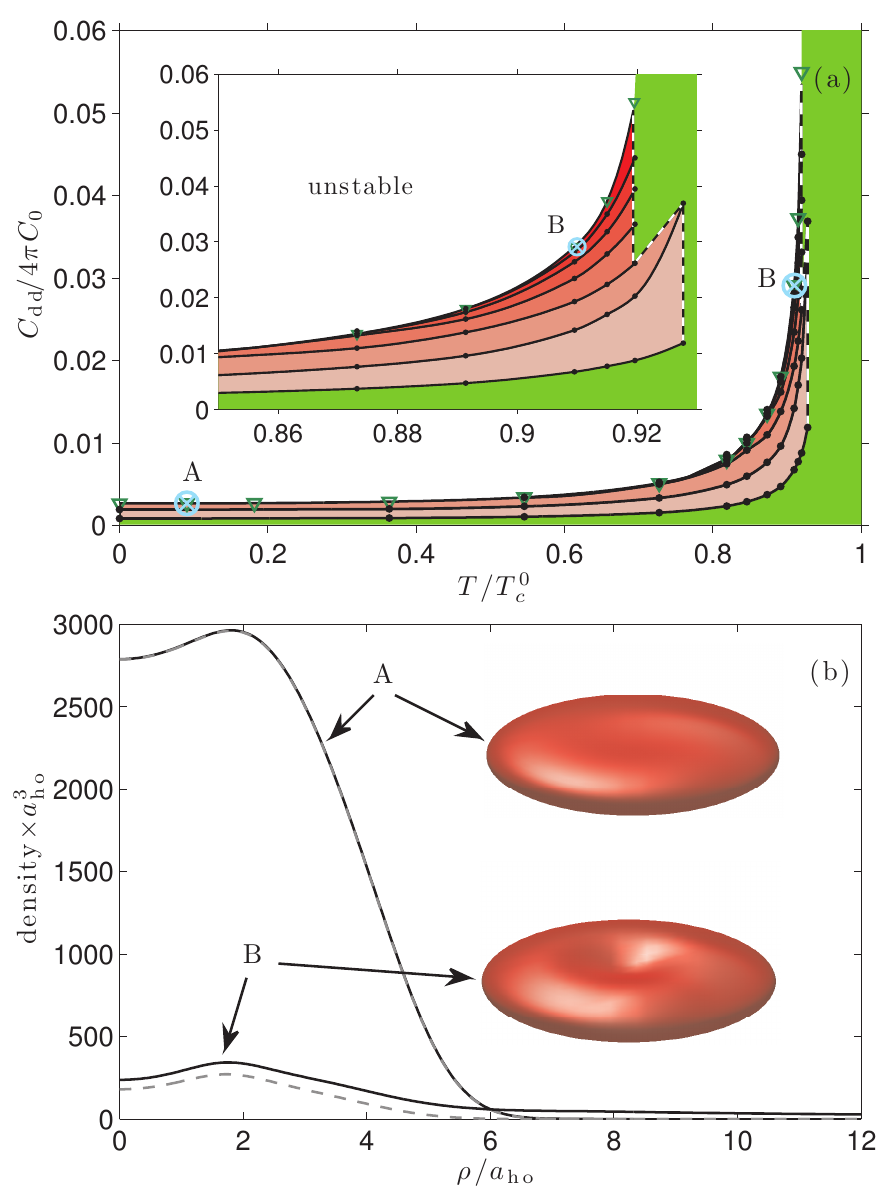}
\caption{(Color online) Biconcave characteristics for $\lambda = 8$ and $N=2\times 10^5$ at finite temperature. (a) Stability diagram with biconcave contrast contours \{0, 0.05, 0.1, 0.15, 0.2, 0.25\} (bottom to top) added. The solid curves are interpolations between the calculated contour points. The white dotted line marks where we terminate the contours due to the condensate fraction becoming negligibly small. Triangles indicate the stability boundary from Fig.~\ref{Fig:StabLog}. Inset: Magnification of the high temperature region.
(b) Radial densities for phase space points marked by A and B in (a). A: $T/T_c^0 = 0.0910$, $C_{\rm{dd}}/ 4\pi C_0 = 0.00268$ and $N_0/N = 1.00$ (thermal depletion $<1\%$). B: $T/T_c^0 =  0.910$, $C_{\rm{dd}}/ 4\pi C_0 = 0.0291$  and $N_0/N = 0.0716$.  Solid (dashed) lines represent the total (condensate) density. Insets: corresponding surface contours at 67\% of the peak density.\\
 \label{Fig:Biconcave}}
\end{center}\vspace{-0.4cm}
\end{figure}

As our final application we consider thermal effects on the biconcave density profiles which occur in oblate geometries. 
To date, the only study of temperature effects of these states was in Ref.~\cite{Ronen2007b} [which we reproduce in Fig.~\ref{Fig:HFBCompare}(b)].
That study considered a single line (at fixed $\Cdd$ and $N$ and varying $T$) through the phase diagram, and showed that biconcavity persisted at small finite temperatures ($T\lesssim0.25T_c^0$), but then was rapidly washed out as temperature  increased further.

Using our Hartree theory we provide a broad characterization of the thermal effects on biconcavity. We focus on the case $\lambda=8$, which supports a biconcave condensate at $T=0$.  In Fig.~\ref{Fig:Biconcave}(a) we present contours of biconcave contrast [as defined in Eq.~(\ref{contrast})] over the entire range of parameters where this state is stable. 
These results show that biconcavity is not destroyed as temperature increases.
Instead the parameter region over which biconcavity occurs grows, with large biconcave contrasts emerging at higher temperature. 
The general trends seen can be understood by considering the thermal depletion of the condensate, using similar arguments to those made to obtain Eq.~(\ref{simplemodel}): as the temperature increases the value of $\Cdd$  required for the condensate to exhibit a biconcave density profile should increase in a manner that is approximately inversely proportional to the condensate occupation.
Thus, the washing out observed in \cite{Ronen2007b} [our Fig.~\ref{Fig:HFBCompare}(b)] arises because they considered $\Cdd$ fixed. 
Thermal depletion of the condensate is not sufficient to explain all aspects observed in our results, e.g.~the deepening of the biconcave contrast that develops at higher temperatures in Fig.~\ref{Fig:Biconcave}(a). This arises from additional effects of the thermal interaction with the condensate, e.g.~small changes in the aspect ratio of the effective potential that the condensate experiences can significantly change the contrast (c.f.~the strong dependence of biconcavity on trap aspect ratio near $\lambda=8$ in Fig.~1 of Ref.~\cite{Ronen2007a}).

 In Fig.~\ref{Fig:Biconcave}(b) we show two examples of the biconcave density profiles at different temperatures. Case B displays the very pronounced biconcavity for a system at $T\approx0.9T_c^0$, where the condensate fraction is $N_0/N\approx0.07$.

\section{Conclusions} 
In this paper we have developed a Hartree theory for a trapped dipolar Bose gas that can be applied to make predictions above and below the condensation temperature.
We have used this theory to quantify the role of thermal fluctuations on the mechanical stability of the cloud, and present results for the stability phase diagram as a function of temperature and aspect ratio.
Above $T_c$  our theory predicts significant corrections to the stability boundary over the equivalent  semiclassical theory. Most notably, the semiclassical theory underestimates the size of the stability region for oblate geometries. Below $T_c$ (but at finite temperature) we find that the stability boundary is well described by the zero temperature result after scaling for the thermally depleted condensate.

We have also studied the role of thermal fluctuations on biconcave condensate states.  Our results show that as temperature increases, and the condensate thermally depletes,   the range of interaction parameters  in which these kinds of states can be found  increases.
Furthermore, a large thermal cloud may actually enhance the biconcave contrast making direct imaging of an \textit{in situ} density oscillating state more feasible, see Fig.~\ref{Fig:Biconcave}(b).

An important step for future theoretical studies in the finite temperature regime is to fully include thermal exchange effects. Because a large number of modes   are important for temperatures of the order of $T_c$, full Hartree-Fock calculations will probably not be feasible. It is possible to include exchange interactions semiclassically (c.f.~Fermi studies \cite{Zhang2010a,Baillie2010b}), although our work here has shown beyond semiclassical effects are important even above $T_c$. An interesting possibility is the extension of  classical field methods to thermal dipolar gases (e.g.~\cite{cfieldRev2008,Sau2009a,Blakie2009e,Tomasz2010a}) which may provide a comprehensive description for temperatures around the condensation transition.

 \section*{Acknowledgments}  This work was supported by the Marsden Fund of New Zealand (contract UOO0924).
 
 \appendix
\section{Hartree and Hartree-Fock theory for dipolar Bose gases}\label{sec:appHHF}
In this appendix we describe the full Hartree-Fock theory for the dipolar Bose gas and discuss the reduction to the Hartree form we employ here. We then introduce the semiclassical Hartree approach we use to calculate high energy modes, which are insensitive to beyond-semiclassical effects.

\subsection{Hartree-Fock equations}
The Hartree-Fock theory for a Bose  gas is well-established \cite{BlaizotRipka}, particularly for the case of contact interactions (e.g.~see Refs.~\cite{Houbiers1996,Bergeman1997}). Here we present this theory for a system interacting with a DDI and consider the cases of above and below $T_c$ separately.
 
\subsubsection{Above $T_c$}
The  Hartree-Fock equation for the modes of an uncondensed dipolar Bose gas is 
\begin{align}  
\epsilon_ju_j(\x)&=H_0u_j(\x) +\underbrace{\int d^3\x'\,U_{\mathrm{dd}}(\x'-\x)n(\x')u_j(\x)}_{\text{Hartree/Direct interaction term}}\nonumber \\
&\,+\underbrace{\int d^3\x'\,U_{\mathrm{dd}}(\x'-\x)\tilde {G}(\x,\x')u_j(\x')}_{\text{Fock/Exchange interaction term}}\, ,\label{HFujaboveTc}
\end{align}
where $H_0$ is the single particle Hamiltonian and
\begin{align}
\tilde{G}(\x,\x') &= \sum_jN_j u_j^*(\x')u_j(\x),\\
n(\x) &= \tilde{G}(\x,\x), 
\end{align}
are the first order coherence function and total density, respectively. 

\subsubsection{Below $T_c$}
Below $T_c$ an appreciable number of atoms condense into the lowest energy single particle mode, which we denote as the condensate mode $u_0(\x)$ with respective energy $\epsilon_0$ and occupation $N_0\sim O(N)$. In this regime the Hartree-Fock equations take the form
\begin{align}  
\epsilon_ju_j(\x)&\!=\!H_0u_j(\x)\!+\!\underbrace{\int d^3\x'\,U_{\mathrm{dd}}(\x'-\x)n(\x')u_j(\x)}_{\text{condensate + thermal direct}}\label{HFuj}\\
&\,+\underbrace{\int d^3\x'\,U_{\mathrm{dd}}(\x'-\x)\tilde {G}(\x,\x')u_j(\x')}_{\text{thermal exchange}}\, \nonumber\\
&\,+\underbrace{Q\left\{\int d^3\x'\,U_{\mathrm{dd}}(\x'-\x)G_{0}(\x',\x) Q\left\{u_j(\x')\right\}\right\}}_{\text{condensate exchange}},\, \nonumber 
\end{align}
 where
\begin{align}
G_{0}(\x',\x) &= N_0u_0^*(\x')u_0(\x),\\
\tilde{G}(\x',\x) &= \sum_{j>0}N_j u_j^*(\x')u_j(\x), \\
n(\x) &= G_{0}(\x,\x)+\tilde{G}(\x,\x) , 
\end{align} 
are the condensate and thermal first order coherence functions, and the total density, respectively.  
We have also introduced the projector
\begin{equation}
Q\{f(\x)\}\equiv \int d^3\mathbf{y}\,\left[\delta(\x-\mathbf{y})-u_0(\x)u_0^*(\mathbf{y})\right]f(\mathbf{y}),
\end{equation} 
to remove components of $f(\x)$ parallel to the condensate mode. The projection operator in Eq.~(\ref{HFuj}) acts to ensure that atoms within the condensate do not undergo an exchange interaction with themselves. In particular, when acting on the condensate mode Eq.~(\ref{HFuj}) reduces to the expected generalized Gross-Pitaevskii equation
\begin{align}
\epsilon_0u_0(\x)&=\left[H_0 +\int d^3\x'\,U_{\mathrm{dd}}(\x'-\x)n(\x')\right]u_0(\x)\label{HFu0}\\
&\,+ {\int d^3\x'\,U_{\mathrm{dd}}(\x'-\x)\tilde {G}(\x',\x)u_0(\x')} ,\, \nonumber 
\end{align}
which has direct interactions with condensate and thermal atoms, but only thermal exchange.
The projector also ensures that the modes $\{u_j(\x)\}$ form an orthonormal set (e.g.~see \cite{Castin1998a,Morgan2000a}) \footnote{Note in Bogoliubov theory many practitioners neglect to perform the projection, which fortuitously does not change the quasiparticle energies. However, in Hartree-Fock theory the mode energies are affected by projection.}.

\subsection{Reduction to Hartree theory} 
The full numerical solution of Eq.~(\ref{HFuj}) [or even Eq.~(\ref{HFujaboveTc})] is extremely challenging, because the thermal exchange requires $\tilde{G}(\x',\x)$ to be calculated (or at least applied) for each self-consistent iteration. This limits the theory to applications involving a small number of modes and away from regimes where mechanical stability can be studied.

 The Hartree theory we use is obtained by neglecting both condensate and thermal exchange terms  [as labeled in Eqs.~(\ref{HFuj})], yielding
\begin{align}  
\epsilon_ju_j(\x)&=\left[H_0 +\int d^3\x'\,U_{\mathrm{dd}}(\x'-\x)n(\x')\right]u_j(\x) \label{Huj},
\end{align}
[i.e.~Eq.~\ref{Hu}].
The absence of exchange terms means that a projector is no longer needed. 

The properties of the Hartree, Hartree-Fock and other theories for the homogeneous
Bose condensed gas, including conservation laws, are extensively discussed in Sec.
VI Ref. \cite{Hohenberg1965a} (also see \cite{TheoryA6} for a discussion of the
HFB and HFBP theories of the inhomogeneous system). We note that in the uniform
purely dipolar gas, the  Hartree term is zero and the DDIs affect the system
only though the Fock term. However, in the trapped system the Hartree term is often
dominant, particularly when the harmonic trap is appreciably anisotropic (i.e.
$\lambda\gg1$ or $\lambda\ll1$).

\section{Description of Hartree algorithm}\label{sec:appnum}
In this section we discuss our implementation of the Hartree model as a numerical algorithm. 
\subsection{Semi-classical treatment of high energy modes}
\begin{figure}[!htbp] 
   \centering
   \includegraphics[width=3.2in]{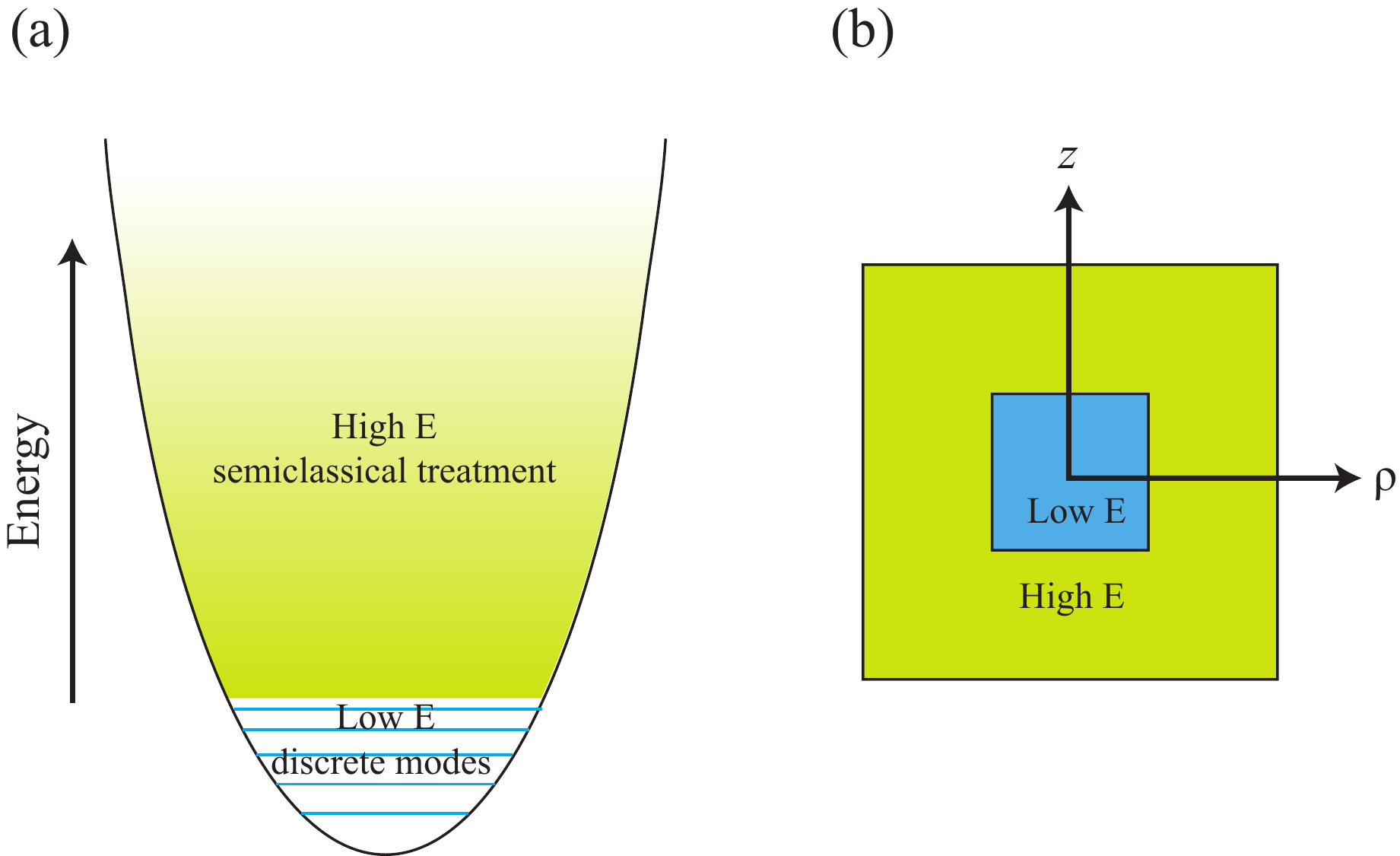} 
   \caption{(Color online) Schematic of our approach to solving dipolar Hartree equations. (a) The low energy (discrete) modes are explicitly solved for by diagonalizing the Hartree equations, whereas higher energy modes are treated using a semiclassical approximation. (b) Schematic of the range of the spatial grids used to solve for the discrete modes and the large grid used to solve for the high energy semiclassical theory. }
   \label{fig:schematic}
\end{figure}

The Hartree equation (\ref{Hu}) is cylindrically symmetric and can be solved using a set of two-dimensional grids. However, at finite temperature typically $\gtrsim10^5$ modes are thermally accessible in the regimes we study, and a full self-consistent calculation is not feasible in terms of the discrete (i.e.~diagonalized) modes. Instead we employ a hybrid method and diagonalize for the lowest energy modes, up to some energy $\ec$, and then calculate the remainder within the semiclassical approach [see Fig.~\ref{fig:schematic}(a)].

The semiclassical approach can be obtained by making the replacement $\nabla\to ik$ in Eq.~(\ref{Hu}), where $k$ is a wavevector. This transforms the Hartree equation to an algebraic equation in which the energy is given in ($\x,\bk)$-phase-space  as
\begin{equation}
\epsilon(\x,\bk)=\frac{\hbar^2\bk^2}{2M}+V_{\textrm{eff}}(\x),\label{Esc}
\end{equation}
where  the  {effective potential} was defined in Eq.~(\ref{Veff}). 
The semiclassical portion of the system is described by a (Wigner) distribution function $W(\x,\bk)=\{e^{\beta[\epsilon(\x,\bk)-\mu]}-1\}^{-1}$. For consistency the semiclassical description can only  be applied to regions of  phase-space where $\epsilon(\x,\bk)>\ec$ to avoid double counting of modes. From this we obtain the semiclassical region density
\begin{align}
n_{\mathrm{sc}}(\x)&=\int_{\epsilon>\ec}\frac{d^3\mathbf{k}}{(2\pi)^3}W(\mathbf{x},\mathbf{k}),\\
&=\frac{1}{\lambda_{\mathrm{dB}}^3}\zeta_{3/2}\left(e^{\beta[\mu-V_{\textrm{eff}}(\x)]},\beta K_{\min}(\x)\right),
\end{align}
where $\zeta_{3/2}(z,y)=(2/\sqrt{\pi})\int_{y}^{\infty}(e^t/z-1)^{-1}\sqrt{t}\,dt$ is the incomplete Bose function and
\begin{equation}
K_{\min}(\x)\equiv \max\{\ec-V_{\textrm{eff}}(\x),0\}.
\end{equation} 

\subsection{Summary of algorithm and numerical considerations}

 All the  excitations up to a  given energy $\ec$ are solved for using the Arnoldi algorithm. The associated discrete mode density is constructed
\begin{equation}
n_{\mathrm{d}}(\x)=\sum_{\epsilon_j<\ec}N_j|u_j(\x)|^2,
\end{equation}
  The semiclassical  and the total densities are then evaluated 
\begin{equation}
n(\x)=n_{\mathrm{d}}(\x)+n_{\mathrm{sc}}(\x).
\end{equation} 
  We use fixed point iteration of these steps to ensure self consistency. To avoid oscillations only a small amount of the new prediction for the total density  $(n_{\mathrm{new}})$ is mixed in with the existing prediction $(n_{\mathrm{old}})$, i.e.
  \begin{equation}
  n(\x)\to\lambda_{\mathrm{mix}} n_{\mathrm{new}}(\x)+(1-\lambda_{\mathrm{mix}})n_{\mathrm{old}}(\x),
  \end{equation}
   where $\lambda_{\mathrm{mix}}$ is the mixing parameter. Upon obtaining a self-consistent solution external parameters are adjusted to tune the solutions to a desired macrostate  (e.g.~an outer loop of $\mu$ being adjusted to obtain the correct total number $N$).

We briefly mention a number of  aspects of our algorithm:
\begin{enumerate}
\item We make use of Fourier-Hankel techniques \cite{Ronen2006a} to utilize the cylindrical symmetry and reduce the eigenvalue problem to being two-dimensional. The Fourier-Hankel approach is useful because it allows accurate Fourier transforms  to simplify the evaluation of the convolution required to construct the direct dipolar interaction.  However, the radial Hankel transform  requires a different radial grid for each angular momentum projection quantum number $m$, thus the problem requires a set of two-dimensional grids (we typically diagonalize modes with $m$ up to $10$, i.e.~requiring 11 grids -- generally even more in oblate geometries). This requires extensive transformation of quantities [e.g.~$n(\x)$] between the grids.
 
\item  We use a \textit{cutoff dipole interaction potential} for improved accuracy and to reduce the size of the numerical grids needed. The cutoff potential minimizes interaction between aliased copies of the system (problematic with Fourier methods used for systems with long-range interactions). We make use of both the cylindrical cutoff suggested in \cite{Lu2010a} and the spherical version developed in \cite{Ronen2006a}.

\item We use two grid extents as schematically shown in Fig.~\ref{fig:schematic}(b). Since we only calculate the discrete modes  up to some relatively small energy, $\ec$, we can use a small and dense set of grids to accurately perform the diagonalizations [and obtain $n_{\mathrm{d}}(\x)$]. Then a much larger grid is used for the semiclassical region which extends out to much higher energy, as needed to accurately capture the thermal tails of the system. 

\item In application to mechanical stability, finding self-consistent solutions is challenging and care needs to be taken to ensure that metastable states are not lost prematurely and that a coarse grid does not disguise instability.  In using fixed point iteration effectively, we employ an algorithm to efficiently increase or decrease the mixing speed $\lambda_{\mathrm{mix}}$ during the self-consistent process. We have observed that if $\lambda_{\mathrm{mix}}$ is too large early on in self-consistency iterations a metastable solution may be lost. Normally we start with $\lambda_{\mathrm{mix}}=0.3$  ($\sim0.01$ for biconcave density regions) and appropriately increase or decrease this during the search for a self-consistent solutions depending upon conditions. We also note that care needs to be taken to reliably detect mechanical instability collapse. We perform a number of tests to determine instability including detecting the development of density spikes and large gaps in the low energy spectrum. We have confirmed that these are good signatures of the grid dependent numerical collapse discussed in Sec.~\ref{sec:locateinstability}.
 \end{enumerate}
 
\subsection{Convergence tests of stability boundary}\label{Sec:ErrorBars}
In oblate geometries the self consistent calculations above $T_c$ become increasingly difficult to perform as $\lambda$ increases. For this reason we have not extended our calculations beyond $\lambda=8$. 
The origin of the difficulty is two fold: (i) the effective potential flattens considerably which increases the low energy density of states, meaning that a large number of modes exist below $\ec$. (ii) In oblate geometries the interaction strengths at instability are much larger, and these numerous low energy modes interact strongly with each other.

The important convergence test is that our results are independent of the boundary ($\ec$) separating the discrete low energy modes from the continuous semiclassical spectrum.
The typical error bars shown in  Figs.~\ref{Fig:Stab} and \ref{Fig:StabLog} represent the variation in the stability boundary as $\ec$ was varied (1 $\sigma$ spread obtained from tests where the number of discrete modes ranged over $\sim$ 10 to 1000).
 It is not computationally feasible for us to take $\ec$ high enough for $\lambda\ge4$ (above $T_c$) to get a self-consistent result fully independent of $\ec$.
However, note we do not observe a systematic drift nor monotonic relationship between $\ec$ and the stability boundary.

Below $T_c$ the error bars are instead determined arbitrarily by the bisection tolerance of parameters $\mu$ and $\Cdd$.

\bibliographystyle{apsrev4-1}


\begin{thebibliography}{58}%
\makeatletter
\providecommand \@ifxundefined [1]{%
 \@ifx{#1\undefined}
}%
\providecommand \@ifnum [1]{%
 \ifnum #1\expandafter \@firstoftwo
 \else \expandafter \@secondoftwo
 \fi
}%
\providecommand \@ifx [1]{%
 \ifx #1\expandafter \@firstoftwo
 \else \expandafter \@secondoftwo
 \fi
}%
\providecommand \natexlab [1]{#1}%
\providecommand \enquote  [1]{``#1''}%
\providecommand \bibnamefont  [1]{#1}%
\providecommand \bibfnamefont [1]{#1}%
\providecommand \citenamefont [1]{#1}%
\providecommand \href@noop [0]{\@secondoftwo}%
\providecommand \href [0]{\begingroup \@sanitize@url \@href}%
\providecommand \@href[1]{\@@startlink{#1}\@@href}%
\providecommand \@@href[1]{\endgroup#1\@@endlink}%
\providecommand \@sanitize@url [0]{\catcode `\\12\catcode `\$12\catcode
  `\&12\catcode `\#12\catcode `\^12\catcode `\_12\catcode `\%12\relax}%
\providecommand \@@startlink[1]{}%
\providecommand \@@endlink[0]{}%
\providecommand \url  [0]{\begingroup\@sanitize@url \@url }%
\providecommand \@url [1]{\endgroup\@href {#1}{\urlprefix }}%
\providecommand \urlprefix  [0]{URL }%
\providecommand \Eprint [0]{\href }%
\providecommand \doibase [0]{http://dx.doi.org/}%
\providecommand \selectlanguage [0]{\@gobble}%
\providecommand \bibinfo  [0]{\@secondoftwo}%
\providecommand \bibfield  [0]{\@secondoftwo}%
\providecommand \translation [1]{[#1]}%
\providecommand \BibitemOpen [0]{}%
\providecommand \bibitemStop [0]{}%
\providecommand \bibitemNoStop [0]{.\EOS\space}%
\providecommand \EOS [0]{\spacefactor3000\relax}%
\providecommand \BibitemShut  [1]{\csname bibitem#1\endcsname}%
\let\auto@bib@innerbib\@empty
\bibitem [{\citenamefont {Lahaye}\ \emph {et~al.}(2007)\citenamefont {Lahaye},
  \citenamefont {Menotti}, \citenamefont {Santos}, \citenamefont {Lewenstein},
  \ and\ \citenamefont {Pfau}}]{Lahaye2007a}%
  \BibitemOpen
  \bibfield  {author} {\bibinfo {author} {\bibfnamefont {T.}~\bibnamefont
  {Lahaye}}, \bibinfo {author} {\bibfnamefont {C.}~\bibnamefont {Menotti}},
  \bibinfo {author} {\bibfnamefont {L.}~\bibnamefont {Santos}}, \bibinfo
  {author} {\bibfnamefont {M.}~\bibnamefont {Lewenstein}}, \ and\ \bibinfo {author} {\bibfnamefont
  {T.}~\bibnamefont {Pfau}},\ }\href@noop {} {\bibfield  {journal} {\bibinfo
  {journal} {Rep. Prog. Phys.}\ }\textbf {\bibinfo {volume} {72}},\ \bibinfo {pages}
  {126401} (\bibinfo {year} {2009})}\BibitemShut {NoStop}%
\bibitem [{\citenamefont {Baranov}\ \emph {et~al.}(2002)\citenamefont
  {Baranov}, \citenamefont {Mar\char39{}enko}, \citenamefont {Rychkov},\ and\
  \citenamefont {Shlyapnikov}}]{Baranov2002a}%
  \BibitemOpen
  \bibfield  {author} {\bibinfo {author} {\bibfnamefont {M.~A.}\ \bibnamefont
  {Baranov}}, \bibinfo {author} {\bibfnamefont {M.~S.}\ \bibnamefont
  {Mar\char39{}enko}}, \bibinfo {author} {\bibfnamefont {V.~S.}\ \bibnamefont
  {Rychkov}}, \ and\ \bibinfo {author} {\bibfnamefont {G.~V.}\ \bibnamefont
  {Shlyapnikov}},\ }\href {\doibase 10.1103/PhysRevA.66.013606} {\bibfield
  {journal} {\bibinfo  {journal} {Phys. Rev. A}\ }\textbf {\bibinfo {volume}
  {66}},\ \bibinfo {pages} {013606} (\bibinfo {year} {2002})}\BibitemShut
  {NoStop}%
\bibitem [{\citenamefont {G\'oral}\ \emph {et~al.}(2002)\citenamefont
  {G\'oral}, \citenamefont {Santos},\ and\ \citenamefont
  {Lewenstein}}]{Goral2002a}%
  \BibitemOpen
  \bibfield  {author} {\bibinfo {author} {\bibfnamefont {K.}~\bibnamefont
  {G\'oral}}, \bibinfo {author} {\bibfnamefont {L.}~\bibnamefont {Santos}}, \
  and\ \bibinfo {author} {\bibfnamefont {M.}~\bibnamefont {Lewenstein}},\
  }\href {\doibase 10.1103/PhysRevLett.88.170406} {\bibfield  {journal}
  {\bibinfo  {journal} {Phys. Rev. Lett.}\ }\textbf {\bibinfo {volume} {88}},\
  \bibinfo {pages} {170406} (\bibinfo {year} {2002})}\BibitemShut {NoStop}%
\bibitem [{\citenamefont {DeMille}(2002)}]{DeMille2002a}%
  \BibitemOpen
  \bibfield  {author} {\bibinfo {author} {\bibfnamefont {D.}~\bibnamefont
  {DeMille}},\ }\href {\doibase 10.1103/PhysRevLett.88.067901} {\bibfield
  {journal} {\bibinfo  {journal} {Phys. Rev. Lett.}\ }\textbf {\bibinfo
  {volume} {88}},\ \bibinfo {pages} {067901} (\bibinfo {year}
  {2002})}\BibitemShut {NoStop}%
\bibitem [{\citenamefont {O\char39{}Dell}\ \emph {et~al.}(2003)\citenamefont
  {O\char39{}Dell}, \citenamefont {Giovanazzi},\ and\ \citenamefont
  {Kurizki}}]{ODell2003a}%
  \BibitemOpen
  \bibfield  {author} {\bibinfo {author} {\bibfnamefont {D.~H.~J.}\
  \bibnamefont {O\char39{}Dell}}, \bibinfo {author} {\bibfnamefont
  {S.}~\bibnamefont {Giovanazzi}}, \ and\ \bibinfo {author} {\bibfnamefont
  {G.}~\bibnamefont {Kurizki}},\ }\href {\doibase
  10.1103/PhysRevLett.90.110402} {\bibfield  {journal} {\bibinfo  {journal}
  {Phys. Rev. Lett.}\ }\textbf {\bibinfo {volume} {90}},\ \bibinfo {pages}
  {110402} (\bibinfo {year} {2003})}\BibitemShut {NoStop}%
\bibitem [{\citenamefont {Kawaguchi}\ \emph {et~al.}(2006)\citenamefont
  {Kawaguchi}, \citenamefont {Saito},\ and\ \citenamefont
  {Ueda}}]{Kawaguchi2006a}%
  \BibitemOpen
  \bibfield  {author} {\bibinfo {author} {\bibfnamefont {Y.}~\bibnamefont
  {Kawaguchi}}, \bibinfo {author} {\bibfnamefont {H.}~\bibnamefont {Saito}}, \
  and\ \bibinfo {author} {\bibfnamefont {M.}~\bibnamefont {Ueda}},\ }\href
  {\doibase 10.1103/PhysRevLett.96.080405} {\bibfield  {journal} {\bibinfo
  {journal} {Phys. Rev. Lett.}\ }\textbf {\bibinfo {volume} {96}},\ \bibinfo
  {eid} {080405} (\bibinfo {year} {2006})}\BibitemShut {NoStop}%
\bibitem [{\citenamefont {Rabl}\ \emph {et~al.}(2006)\citenamefont {Rabl},
  \citenamefont {DeMille}, \citenamefont {Doyle}, \citenamefont {Lukin},
  \citenamefont {Schoelkopf},\ and\ \citenamefont {Zoller}}]{Rabl2006a}%
  \BibitemOpen
  \bibfield  {author} {\bibinfo {author} {\bibfnamefont {P.}~\bibnamefont
  {Rabl}}, \bibinfo {author} {\bibfnamefont {D.}~\bibnamefont {DeMille}},
  \bibinfo {author} {\bibfnamefont {J.~M.}\ \bibnamefont {Doyle}}, \bibinfo
  {author} {\bibfnamefont {M.~D.}\ \bibnamefont {Lukin}}, \bibinfo {author}
  {\bibfnamefont {R.~J.}\ \bibnamefont {Schoelkopf}}, \ and\ \bibinfo {author}
  {\bibfnamefont {P.}~\bibnamefont {Zoller}},\ }\href {\doibase
  10.1103/PhysRevLett.97.033003} {\bibfield  {journal} {\bibinfo  {journal}
  {Phys. Rev. Lett.}\ }\textbf {\bibinfo {volume} {97}},\ \bibinfo {eid}
  {033003} (\bibinfo {year} {2006})}\BibitemShut {NoStop}%
\bibitem [{\citenamefont {B\"{u}chler}\ \emph {et~al.}(2007)\citenamefont
  {B\"{u}chler}, \citenamefont {Demler}, \citenamefont {Lukin}, \citenamefont
  {Micheli}, \citenamefont {Prokof'ev}, \citenamefont {Pupillo},\ and\
  \citenamefont {Zoller}}]{Buchler2007a}%
  \BibitemOpen
  \bibfield  {author} {\bibinfo {author} {\bibfnamefont {H.~P.}\ \bibnamefont
  {B\"{u}chler}}, \bibinfo {author} {\bibfnamefont {E.}~\bibnamefont {Demler}},
  \bibinfo {author} {\bibfnamefont {M.}~\bibnamefont {Lukin}}, \bibinfo
  {author} {\bibfnamefont {A.}~\bibnamefont {Micheli}}, \bibinfo {author}
  {\bibfnamefont {N.}~\bibnamefont {Prokof'ev}}, \bibinfo {author}
  {\bibfnamefont {G.}~\bibnamefont {Pupillo}}, \ and\ \bibinfo {author}
  {\bibfnamefont {P.}~\bibnamefont {Zoller}},\ }\href {\doibase
  10.1103/PhysRevLett.98.060404} {\bibfield  {journal} {\bibinfo  {journal}
  {Phys. Rev. Lett.}\ }\textbf {\bibinfo {volume} {98}},\ \bibinfo {eid}
  {060404} (\bibinfo {year} {2007})}\BibitemShut {NoStop}%
\bibitem [{\citenamefont {Griesmaier}\ \emph {et~al.}(2005)\citenamefont
  {Griesmaier}, \citenamefont {Werner}, \citenamefont {Hensler}, \citenamefont
  {Stuhler},\ and\ \citenamefont {Pfau}}]{Griesmaier2005a}%
  \BibitemOpen
  \bibfield  {author} {\bibinfo {author} {\bibfnamefont {A.}~\bibnamefont
  {Griesmaier}}, \bibinfo {author} {\bibfnamefont {J.}~\bibnamefont {Werner}},
  \bibinfo {author} {\bibfnamefont {S.}~\bibnamefont {Hensler}}, \bibinfo
  {author} {\bibfnamefont {J.}~\bibnamefont {Stuhler}}, \ and\ \bibinfo
  {author} {\bibfnamefont {T.}~\bibnamefont {Pfau}},\ }\href {\doibase
  10.1103/PhysRevLett.94.160401} {\bibfield  {journal} {\bibinfo  {journal}
  {Phys. Rev. Lett.}\ }\textbf {\bibinfo {volume} {94}},\ \bibinfo {pages}
  {160401} (\bibinfo {year} {2005})}\BibitemShut {NoStop}%
\bibitem [{\citenamefont {Lahaye}\ \emph {et~al.}(2008)\citenamefont {Lahaye},
  \citenamefont {Metz}, \citenamefont {Fr\"{o}hlich}, \citenamefont {Koch},
  \citenamefont {Meister}, \citenamefont {Griesmaier}, \citenamefont {Pfau},
  \citenamefont {Saito}, \citenamefont {Kawaguchi},\ and\ \citenamefont
  {Ueda}}]{Lahaye2009a}%
  \BibitemOpen
  \bibfield  {author} {\bibinfo {author} {\bibfnamefont {T.}~\bibnamefont
  {Lahaye}}, \bibinfo {author} {\bibfnamefont {J.}~\bibnamefont {Metz}},
  \bibinfo {author} {\bibfnamefont {B.}~\bibnamefont {Fr\"{o}hlich}}, \bibinfo
  {author} {\bibfnamefont {T.}~\bibnamefont {Koch}}, \bibinfo {author}
  {\bibfnamefont {M.}~\bibnamefont {Meister}}, \bibinfo {author} {\bibfnamefont
  {A.}~\bibnamefont {Griesmaier}}, \bibinfo {author} {\bibfnamefont
  {T.}~\bibnamefont {Pfau}}, \bibinfo {author} {\bibfnamefont {H.}~\bibnamefont
  {Saito}}, \bibinfo {author} {\bibfnamefont {Y.}~\bibnamefont {Kawaguchi}}, \
  and\ \bibinfo {author} {\bibfnamefont {M.}~\bibnamefont {Ueda}},\ }\href
  {\doibase 10.1103/PhysRevLett.101.080401} {\bibfield  {journal} {\bibinfo
  {journal} {Phys. Rev. Lett.}\ }\textbf {\bibinfo {volume} {101}},\ \bibinfo
  {eid} {080401} (\bibinfo {year} {2008})}\BibitemShut {NoStop}%
\bibitem [{\citenamefont {Bismut}\ \emph {et~al.}(2010)\citenamefont {Bismut},
  \citenamefont {Pasquiou}, \citenamefont {Mar\'echal}, \citenamefont {Pedri},
  \citenamefont {Vernac}, \citenamefont {Gorceix},\ and\ \citenamefont
  {Laburthe-Tolra}}]{Bismut2010a}%
  \BibitemOpen
  \bibfield  {author} {\bibinfo {author} {\bibfnamefont {G.}~\bibnamefont
  {Bismut}}, \bibinfo {author} {\bibfnamefont {B.}~\bibnamefont {Pasquiou}},
  \bibinfo {author} {\bibfnamefont {E.}~\bibnamefont {Mar\'echal}}, \bibinfo
  {author} {\bibfnamefont {P.}~\bibnamefont {Pedri}}, \bibinfo {author}
  {\bibfnamefont {L.}~\bibnamefont {Vernac}}, \bibinfo {author} {\bibfnamefont
  {O.}~\bibnamefont {Gorceix}}, \ and\ \bibinfo {author} {\bibfnamefont
  {B.}~\bibnamefont {Laburthe-Tolra}},\ }\href {\doibase
  10.1103/PhysRevLett.105.040404} {\bibfield  {journal} {\bibinfo  {journal}
  {Phys. Rev. Lett.}\ }\textbf {\bibinfo {volume} {105}},\ \bibinfo {pages}
  {040404} (\bibinfo {year} {2010})}\BibitemShut {NoStop}%
\bibitem [{\citenamefont {Pasquiou}\ \emph {et~al.}(2011)\citenamefont
  {Pasquiou}, \citenamefont {Bismut}, \citenamefont {Mar\'echal}, \citenamefont
  {Pedri}, \citenamefont {Vernac}, \citenamefont {Gorceix},\ and\ \citenamefont
  {Laburthe-Tolra}}]{Pasquiou2011a}%
  \BibitemOpen
  \bibfield  {author} {\bibinfo {author} {\bibfnamefont {B.}~\bibnamefont
  {Pasquiou}}, \bibinfo {author} {\bibfnamefont {G.}~\bibnamefont {Bismut}},
  \bibinfo {author} {\bibfnamefont {E.}~\bibnamefont {Mar\'echal}}, \bibinfo
  {author} {\bibfnamefont {P.}~\bibnamefont {Pedri}}, \bibinfo {author}
  {\bibfnamefont {L.}~\bibnamefont {Vernac}}, \bibinfo {author} {\bibfnamefont
  {O.}~\bibnamefont {Gorceix}}, \ and\ \bibinfo {author} {\bibfnamefont
  {B.}~\bibnamefont {Laburthe-Tolra}},\ }\href {\doibase
  10.1103/PhysRevLett.106.015301} {\bibfield  {journal} {\bibinfo  {journal}
  {Phys. Rev. Lett.}\ }\textbf {\bibinfo {volume} {106}},\ \bibinfo {pages}
  {015301} (\bibinfo {year} {2011})}\BibitemShut {NoStop}%
\bibitem [{\citenamefont {Lu}\ \emph {et~al.}(2011)\citenamefont {Lu},
  \citenamefont {Burdick}, \citenamefont {Youn},\ and\ \citenamefont
  {Lev}}]{Mingwu2011a}%
  \BibitemOpen
  \bibfield  {author} {\bibinfo {author} {\bibfnamefont {M.}~\bibnamefont
  {Lu}}, \bibinfo {author} {\bibfnamefont {N.~Q.}\ \bibnamefont {Burdick}},
  \bibinfo {author} {\bibfnamefont {S.~H.}\ \bibnamefont {Youn}}, \ and\
  \bibinfo {author} {\bibfnamefont {B.~L.}\ \bibnamefont {Lev}},\ }\href
  {\doibase 10.1103/PhysRevLett.107.190401} {\bibfield  {journal} {\bibinfo
  {journal} {Phys. Rev. Lett.}\ }\textbf {\bibinfo {volume} {107}},\ \bibinfo
  {pages} {190401} (\bibinfo {year} {2011})}\BibitemShut {NoStop}%
\bibitem [{\citenamefont {Aikawa}\ \emph {et~al.}(2012)\citenamefont {Aikawa},
  \citenamefont {Frisch}, \citenamefont {Mark}, \citenamefont {Baier},
  \citenamefont {Rietzler}, \citenamefont {Grimm},\ and\ \citenamefont
  {Ferlaino}}]{Aikawa2012a}%
  \BibitemOpen
  \bibfield  {author} {\bibinfo {author} {\bibfnamefont {K.}~\bibnamefont
  {Aikawa}}, \bibinfo {author} {\bibfnamefont {A.}~\bibnamefont {Frisch}},
  \bibinfo {author} {\bibfnamefont {M.}~\bibnamefont {Mark}}, \bibinfo {author}
  {\bibfnamefont {S.}~\bibnamefont {Baier}}, \bibinfo {author} {\bibfnamefont
  {A.}~\bibnamefont {Rietzler}}, \bibinfo {author} {\bibfnamefont
  {R.}~\bibnamefont {Grimm}}, \ and\ \bibinfo {author} {\bibfnamefont
  {F.}~\bibnamefont {Ferlaino}},\ }\href {\doibase
  10.1103/PhysRevLett.108.210401} {\bibfield  {journal} {\bibinfo  {journal}
  {Phys. Rev. Lett.}\ }\textbf {\bibinfo {volume} {108}},\ \bibinfo {pages}
  {210401} (\bibinfo {year} {2012})}\BibitemShut {NoStop}%
\bibitem [{\citenamefont {Aikawa}\ \emph {et~al.}(2010)\citenamefont {Aikawa},
  \citenamefont {Akamatsu}, \citenamefont {Hayashi}, \citenamefont {Oasa},
  \citenamefont {Kobayashi}, \citenamefont {Naidon}, \citenamefont {Kishimoto},
  \citenamefont {Ueda},\ and\ \citenamefont {Inouye}}]{Aikawa2010a}%
  \BibitemOpen
  \bibfield  {author} {\bibinfo {author} {\bibfnamefont {K.}~\bibnamefont
  {Aikawa}}, \bibinfo {author} {\bibfnamefont {D.}~\bibnamefont {Akamatsu}},
  \bibinfo {author} {\bibfnamefont {M.}~\bibnamefont {Hayashi}}, \bibinfo
  {author} {\bibfnamefont {K.}~\bibnamefont {Oasa}}, \bibinfo {author}
  {\bibfnamefont {J.}~\bibnamefont {Kobayashi}}, \bibinfo {author}
  {\bibfnamefont {P.}~\bibnamefont {Naidon}}, \bibinfo {author} {\bibfnamefont
  {T.}~\bibnamefont {Kishimoto}}, \bibinfo {author} {\bibfnamefont
  {M.}~\bibnamefont {Ueda}}, \ and\ \bibinfo {author} {\bibfnamefont
  {S.}~\bibnamefont {Inouye}},\ }\href {\doibase
  10.1103/PhysRevLett.105.203001} {\bibfield  {journal} {\bibinfo  {journal}
  {Phys. Rev. Lett.}\ }\textbf {\bibinfo {volume} {105}},\ \bibinfo {pages}
  {203001} (\bibinfo {year} {2010})}\BibitemShut {NoStop}%
\bibitem [{\citenamefont {Ni}\ \emph {et~al.}(2008)\citenamefont {Ni},
  \citenamefont {Ospelkaus}, \citenamefont {de~Miranda}, \citenamefont {Pe'er},
  \citenamefont {Neyenhuis}, \citenamefont {Zirbel}, \citenamefont
  {Kotochigova}, \citenamefont {Julienne}, \citenamefont {Jin},\ and\
  \citenamefont {Ye}}]{Ni2008a}%
  \BibitemOpen
  \bibfield  {author} {\bibinfo {author} {\bibfnamefont {K.-K.}\ \bibnamefont
  {Ni}}, \bibinfo {author} {\bibfnamefont {S.}~\bibnamefont {Ospelkaus}},
  \bibinfo {author} {\bibfnamefont {M.~H.~G.}\ \bibnamefont {de~Miranda}},
  \bibinfo {author} {\bibfnamefont {A.}~\bibnamefont {Pe'er}}, \bibinfo
  {author} {\bibfnamefont {B.}~\bibnamefont {Neyenhuis}}, \bibinfo {author}
  {\bibfnamefont {J.~J.}\ \bibnamefont {Zirbel}}, \bibinfo {author}
  {\bibfnamefont {S.}~\bibnamefont {Kotochigova}}, \bibinfo {author}
  {\bibfnamefont {P.~S.}\ \bibnamefont {Julienne}}, \bibinfo {author}
  {\bibfnamefont {D.~S.}\ \bibnamefont {Jin}}, \ and\ \bibinfo {author}
  {\bibfnamefont {J.}~\bibnamefont {Ye}},\ }\href@noop {} {\bibfield  {journal}
  {\bibinfo  {journal} {Science}\ }\textbf {\bibinfo {volume} {322}},\ \bibinfo
  {pages} {231} (\bibinfo {year} {2008})}\BibitemShut {NoStop}%
\bibitem [{\citenamefont {Lu}\ \emph {et~al.}(2012)\citenamefont {Lu},
  \citenamefont {Burdick},\ and\ \citenamefont {Lev}}]{Lu2012a}%
  \BibitemOpen
  \bibfield  {author} {\bibinfo {author} {\bibfnamefont {M.}~\bibnamefont
  {Lu}}, \bibinfo {author} {\bibfnamefont {N.~Q.}\ \bibnamefont {Burdick}}, \
  and\ \bibinfo {author} {\bibfnamefont {B.~L.}\ \bibnamefont {Lev}},\ }\href
  {\doibase 10.1103/PhysRevLett.108.215301} {\bibfield  {journal} {\bibinfo
  {journal} {Phys. Rev. Lett.}\ }\textbf {\bibinfo {volume} {108}},\ \bibinfo
  {pages} {215301} (\bibinfo {year} {2012})}\BibitemShut {NoStop}%
\bibitem [{\citenamefont {Yi}\ and\ \citenamefont {You}(2001)}]{Yi2001a}%
  \BibitemOpen
  \bibfield  {author} {\bibinfo {author} {\bibfnamefont {S.}~\bibnamefont
  {Yi}}\ and\ \bibinfo {author} {\bibfnamefont {L.}~\bibnamefont {You}},\
  }\href {\doibase 10.1103/PhysRevA.63.053607} {\bibfield  {journal} {\bibinfo
  {journal} {Phys. Rev. A}\ }\textbf {\bibinfo {volume} {63}},\ \bibinfo
  {pages} {053607} (\bibinfo {year} {2001})}\BibitemShut {NoStop}%
\bibitem [{\citenamefont {Eberlein}\ \emph {et~al.}(2005)\citenamefont
  {Eberlein}, \citenamefont {Giovanazzi},\ and\ \citenamefont
  {O'Dell}}]{Eberlein2005a}%
  \BibitemOpen
  \bibfield  {author} {\bibinfo {author} {\bibfnamefont {C.}~\bibnamefont
  {Eberlein}}, \bibinfo {author} {\bibfnamefont {S.}~\bibnamefont
  {Giovanazzi}}, \ and\ \bibinfo {author} {\bibfnamefont {D.~H.~J.}\
  \bibnamefont {O'Dell}},\ }\href {\doibase 10.1103/PhysRevA.71.033618}
  {\bibfield  {journal} {\bibinfo  {journal} {Phys. Rev. A}\ }\textbf {\bibinfo
  {volume} {71}},\ \bibinfo {pages} {033618} (\bibinfo {year}
  {2005})}\BibitemShut {NoStop}%
\bibitem [{\citenamefont {Ronen}\ \emph {et~al.}(2007)\citenamefont {Ronen},
  \citenamefont {Bortolotti},\ and\ \citenamefont {Bohn}}]{Ronen2007a}%
  \BibitemOpen
  \bibfield  {author} {\bibinfo {author} {\bibfnamefont {S.}~\bibnamefont
  {Ronen}}, \bibinfo {author} {\bibfnamefont {D.~C.~E.}\ \bibnamefont
  {Bortolotti}}, \ and\ \bibinfo {author} {\bibfnamefont {J.~L.}\ \bibnamefont
  {Bohn}},\ }\href {\doibase 10.1103/PhysRevLett.98.030406} {\bibfield
  {journal} {\bibinfo  {journal} {Phys. Rev. Lett.}\ }\textbf {\bibinfo
  {volume} {98}},\ \bibinfo {eid} {030406} (\bibinfo {year}
  {2007})}\BibitemShut {NoStop}%
\bibitem [{\citenamefont {Lu}\ \emph {et~al.}(2010)\citenamefont {Lu},
  \citenamefont {Lu}, \citenamefont {Zhang}, \citenamefont {Qiu}, \citenamefont
  {Pu},\ and\ \citenamefont {Yi}}]{Lu2010a}%
  \BibitemOpen
  \bibfield  {author} {\bibinfo {author} {\bibfnamefont {H.-Y.}\ \bibnamefont
  {Lu}}, \bibinfo {author} {\bibfnamefont {H.}~\bibnamefont {Lu}}, \bibinfo
  {author} {\bibfnamefont {J.-N.}\ \bibnamefont {Zhang}}, \bibinfo {author}
  {\bibfnamefont {R.-Z.}\ \bibnamefont {Qiu}}, \bibinfo {author} {\bibfnamefont
  {H.}~\bibnamefont {Pu}}, \ and\ \bibinfo {author} {\bibfnamefont
  {S.}~\bibnamefont {Yi}},\ }\href {\doibase 10.1103/PhysRevA.82.023622}
  {\bibfield  {journal} {\bibinfo  {journal} {Phys. Rev. A}\ }\textbf {\bibinfo
  {volume} {82}},\ \bibinfo {pages} {023622} (\bibinfo {year}
  {2010})}\BibitemShut {NoStop}%
\bibitem [{\citenamefont {Wilson}\ \emph
  {et~al.}(2009{\natexlab{a}})\citenamefont {Wilson}, \citenamefont {Ronen},\
  and\ \citenamefont {Bohn}}]{Wilson2009a}%
  \BibitemOpen
  \bibfield  {author} {\bibinfo {author} {\bibfnamefont {R.~M.}\ \bibnamefont
  {Wilson}}, \bibinfo {author} {\bibfnamefont {S.}~\bibnamefont {Ronen}}, \
  and\ \bibinfo {author} {\bibfnamefont {J.~L.}\ \bibnamefont {Bohn}},\ }\href
  {\doibase 10.1103/PhysRevA.80.023614} {\bibfield  {journal} {\bibinfo
  {journal} {Phys. Rev. A}\ }\textbf {\bibinfo {volume} {80}},\ \bibinfo
  {pages} {023614} (\bibinfo {year} {2009}{\natexlab{a}})}\BibitemShut
  {NoStop}%
\bibitem [{\citenamefont {Wilson}\ and\ \citenamefont
  {Bohn}(2011)}]{Wilson2011a}%
  \BibitemOpen
  \bibfield  {author} {\bibinfo {author} {\bibfnamefont {R.~M.}\ \bibnamefont
  {Wilson}}\ and\ \bibinfo {author} {\bibfnamefont {J.~L.}\ \bibnamefont
  {Bohn}},\ }\href {\doibase 10.1103/PhysRevA.83.023623} {\bibfield  {journal}
  {\bibinfo  {journal} {Phys. Rev. A}\ }\textbf {\bibinfo {volume} {83}},\
  \bibinfo {pages} {023623} (\bibinfo {year} {2011})}\BibitemShut {NoStop}%
\bibitem [{\citenamefont {Bisset}\ \emph {et~al.}(2011)\citenamefont {Bisset},
  \citenamefont {Baillie},\ and\ \citenamefont {Blakie}}]{Bisset2011}%
  \BibitemOpen
  \bibfield  {author} {\bibinfo {author} {\bibfnamefont {R.~N.}\ \bibnamefont
  {Bisset}}, \bibinfo {author} {\bibfnamefont {D.}~\bibnamefont {Baillie}}, \
  and\ \bibinfo {author} {\bibfnamefont {P.~B.}\ \bibnamefont {Blakie}},\
  }\href {\doibase 10.1103/PhysRevA.83.061602} {\bibfield  {journal} {\bibinfo
  {journal} {Phys. Rev. A}\ }\textbf {\bibinfo {volume} {83}},\ \bibinfo
  {pages} {061602} (\bibinfo {year} {2011})}\BibitemShut {NoStop}%
\bibitem [{\citenamefont {Koch}\ \emph {et~al.}(2008)\citenamefont {Koch},
  \citenamefont {Lahaye}, \citenamefont {Metz}, \citenamefont {B.Froehlich},
  \citenamefont {Griesmaier},\ and\ \citenamefont {Pfau}}]{Koch2008a}%
  \BibitemOpen
  \bibfield  {author} {\bibinfo {author} {\bibfnamefont {T.}~\bibnamefont
  {Koch}}, \bibinfo {author} {\bibfnamefont {T.}~\bibnamefont {Lahaye}},
  \bibinfo {author} {\bibfnamefont {J.}~\bibnamefont {Metz}}, \bibinfo {author}
  {\bibnamefont {B.Froehlich}}, \bibinfo {author} {\bibfnamefont
  {A.}~\bibnamefont {Griesmaier}}, \ and\ \bibinfo {author} {\bibfnamefont
  {T.}~\bibnamefont {Pfau}},\ }\href@noop {} {\bibfield  {journal} {\bibinfo
  {journal} {Nat. Phys.}\ }\textbf {\bibinfo {volume} {4}},\ \bibinfo {pages}
  {218} (\bibinfo {year} {2008})}\BibitemShut {NoStop}%
\bibitem [{\citenamefont {Ronen}\ \emph {et~al.}(2006)\citenamefont {Ronen},
  \citenamefont {Bortolotti},\ and\ \citenamefont {Bohn}}]{Ronen2006a}%
  \BibitemOpen
  \bibfield  {author} {\bibinfo {author} {\bibfnamefont {S.}~\bibnamefont
  {Ronen}}, \bibinfo {author} {\bibfnamefont {D.~C.~E.}\ \bibnamefont
  {Bortolotti}}, \ and\ \bibinfo {author} {\bibfnamefont {J.~L.}\ \bibnamefont
  {Bohn}},\ }\href@noop {} {\bibfield  {journal} {\bibinfo  {journal} {Phys.
  Rev. A}\ }\textbf {\bibinfo {volume} {74}},\ \bibinfo {eid} {013623}
  (\bibinfo {year} {2006})}\BibitemShut {NoStop}%
\bibitem [{\citenamefont {Ronen}\ and\ \citenamefont
  {Bohn}(2007)}]{Ronen2007b}%
  \BibitemOpen
  \bibfield  {author} {\bibinfo {author} {\bibfnamefont {S.}~\bibnamefont
  {Ronen}}\ and\ \bibinfo {author} {\bibfnamefont {J.~L.}\ \bibnamefont
  {Bohn}},\ }\href@noop {} {\bibfield  {journal} {\bibinfo  {journal} {Phys.
  Rev. A}\ }\textbf {\bibinfo {volume} {76}},\ \bibinfo {eid} {043607}
  (\bibinfo {year} {2007})}\BibitemShut {NoStop}%
\bibitem [{\citenamefont {He}\ \emph {et~al.}(2008)\citenamefont {He},
  \citenamefont {Zhang}, \citenamefont {Zhang},\ and\ \citenamefont
  {Yi}}]{He2008a}%
  \BibitemOpen
  \bibfield  {author} {\bibinfo {author} {\bibfnamefont {L.}~\bibnamefont
  {He}}, \bibinfo {author} {\bibfnamefont {J.-N.}\ \bibnamefont {Zhang}},
  \bibinfo {author} {\bibfnamefont {Y.}~\bibnamefont {Zhang}}, \ and\ \bibinfo
  {author} {\bibfnamefont {S.}~\bibnamefont {Yi}},\ }\href {\doibase
  10.1103/PhysRevA.77.031605} {\bibfield  {journal} {\bibinfo  {journal} {Phys.
  Rev. A}\ }\textbf {\bibinfo {volume} {77}},\ \bibinfo {pages} {031605}
  (\bibinfo {year} {2008})}\BibitemShut {NoStop}%
\bibitem [{\citenamefont {Zhang}\ and\ \citenamefont {Yi}(2010)}]{Zhang2010a}%
  \BibitemOpen
  \bibfield  {author} {\bibinfo {author} {\bibfnamefont {J.-N.}\ \bibnamefont
  {Zhang}}\ and\ \bibinfo {author} {\bibfnamefont {S.}~\bibnamefont {Yi}},\
  }\href {\doibase 10.1103/PhysRevA.81.033617} {\bibfield  {journal} {\bibinfo
  {journal} {Phys. Rev. A}\ }\textbf {\bibinfo {volume} {81}},\ \bibinfo
  {pages} {033617} (\bibinfo {year} {2010})}\BibitemShut {NoStop}%
\bibitem [{\citenamefont {Baillie}\ and\ \citenamefont
  {Blakie}(2010)}]{Baillie2010b}%
  \BibitemOpen
  \bibfield  {author} {\bibinfo {author} {\bibfnamefont {D.}~\bibnamefont
  {Baillie}}\ and\ \bibinfo {author} {\bibfnamefont {P.~B.}\ \bibnamefont
  {Blakie}},\ }\href {\doibase 10.1103/PhysRevA.82.033605} {\bibfield
  {journal} {\bibinfo  {journal} {Phys. Rev. A}\ }\textbf {\bibinfo {volume}
  {82}},\ \bibinfo {pages} {033605} (\bibinfo {year} {2010})}\BibitemShut
  {NoStop}%
\bibitem [{\citenamefont {Ticknor}(2012)}]{Ticknor2012a}%
  \BibitemOpen
  \bibfield  {author} {\bibinfo {author} {\bibfnamefont {C.}~\bibnamefont
  {Ticknor}},\ }\href {\doibase 10.1103/PhysRevA.85.033629} {\bibfield
  {journal} {\bibinfo  {journal} {Phys. Rev. A}\ }\textbf {\bibinfo {volume}
  {85}},\ \bibinfo {pages} {033629} (\bibinfo {year} {2012})}\BibitemShut
  {NoStop}%
\bibitem [{Note1()}]{Note1}%
  \BibitemOpen
  \bibinfo {note} {See Appendix \ref {sec:appHHF} for definitions of these
  exchange terms.}\BibitemShut {Stop}%
\bibitem [{\citenamefont {Dalfovo}\ \emph {et~al.}(1997)\citenamefont
  {Dalfovo}, \citenamefont {Giorgini}, \citenamefont {Guilleumas},
  \citenamefont {Pitaevskii},\ and\ \citenamefont {Stringari}}]{Dalfovo1997a}%
  \BibitemOpen
  \bibfield  {author} {\bibinfo {author} {\bibfnamefont {F.}~\bibnamefont
  {Dalfovo}}, \bibinfo {author} {\bibfnamefont {S.}~\bibnamefont {Giorgini}},
  \bibinfo {author} {\bibfnamefont {M.}~\bibnamefont {Guilleumas}}, \bibinfo
  {author} {\bibfnamefont {L.}~\bibnamefont {Pitaevskii}}, \ and\ \bibinfo
  {author} {\bibfnamefont {S.}~\bibnamefont {Stringari}},\ }\href {\doibase
  10.1103/PhysRevA.56.3840} {\bibfield  {journal} {\bibinfo  {journal} {Phys.
  Rev. A}\ }\textbf {\bibinfo {volume} {56}},\ \bibinfo {pages} {3840}
  (\bibinfo {year} {1997})}\BibitemShut {NoStop}%
\bibitem [{\citenamefont {Houbiers}\ and\ \citenamefont
  {Stoof}(1996)}]{Houbiers1996}%
  \BibitemOpen
  \bibfield  {author} {\bibinfo {author} {\bibfnamefont {M.}~\bibnamefont
  {Houbiers}}\ and\ \bibinfo {author} {\bibfnamefont {H.~T.~C.}\ \bibnamefont
  {Stoof}},\ }\href {\doibase 10.1103/PhysRevA.54.5055} {\bibfield  {journal}
  {\bibinfo  {journal} {Phys. Rev. A}\ }\textbf {\bibinfo {volume} {54}},\
  \bibinfo {pages} {5055} (\bibinfo {year} {1996})}\BibitemShut {NoStop}%
\bibitem [{\citenamefont {Bergeman}(1997)}]{Bergeman1997}%
  \BibitemOpen
  \bibfield  {author} {\bibinfo {author} {\bibfnamefont {T.}~\bibnamefont
  {Bergeman}},\ }\href {\doibase 10.1103/PhysRevA.55.3658} {\bibfield
  {journal} {\bibinfo  {journal} {Phys. Rev. A}\ }\textbf {\bibinfo {volume}
  {55}},\ \bibinfo {pages} {3658} (\bibinfo {year} {1997})}\BibitemShut
  {NoStop}%
\bibitem [{\citenamefont {Davis}\ \emph {et~al.}(1999)\citenamefont {Davis},
  \citenamefont {Hutchinson},\ and\ \citenamefont {Zaremba}}]{Davis1999}%
  \BibitemOpen
  \bibfield  {author} {\bibinfo {author} {\bibfnamefont {M.~J.}\ \bibnamefont
  {Davis}}, \bibinfo {author} {\bibfnamefont {D.~A.~W.}\ \bibnamefont
  {Hutchinson}}, \ and\ \bibinfo {author} {\bibfnamefont {E.}~\bibnamefont
  {Zaremba}},\ }\href {http://stacks.iop.org/0953-4075/32/i=15/a=327}
  {\bibfield  {journal} {\bibinfo  {journal} {Journal of Physics B: Atomic,
  Molecular and Optical Physics}\ }\textbf {\bibinfo {volume} {32}},\ \bibinfo
  {pages} {3993} (\bibinfo {year} {1999})}\BibitemShut {NoStop}%
\bibitem [{\citenamefont {Bradley}\ \emph {et~al.}(1995)\citenamefont
  {Bradley}, \citenamefont {Sackett}, \citenamefont {Tollett},\ and\
  \citenamefont {Hulet}}]{Bradley1995a}%
  \BibitemOpen
  \bibfield  {author} {\bibinfo {author} {\bibfnamefont {C.~C.}\ \bibnamefont
  {Bradley}}, \bibinfo {author} {\bibfnamefont {C.~A.}\ \bibnamefont
  {Sackett}}, \bibinfo {author} {\bibfnamefont {J.~J.}\ \bibnamefont
  {Tollett}}, \ and\ \bibinfo {author} {\bibfnamefont {R.~G.}\ \bibnamefont
  {Hulet}},\ }\href {\doibase 10.1103/PhysRevLett.75.1687} {\bibfield
  {journal} {\bibinfo  {journal} {Phys. Rev. Lett.}\ }\textbf {\bibinfo
  {volume} {75}},\ \bibinfo {pages} {1687} (\bibinfo {year}
  {1995})}\BibitemShut {NoStop}%
\bibitem [{\citenamefont {Wilson}\ \emph
  {et~al.}(2009{\natexlab{b}})\citenamefont {Wilson}, \citenamefont {Ronen},\
  and\ \citenamefont {Bohn}}]{Wilson2009}%
  \BibitemOpen
  \bibfield  {author} {\bibinfo {author} {\bibfnamefont {R.~M.}\ \bibnamefont
  {Wilson}}, \bibinfo {author} {\bibfnamefont {S.}~\bibnamefont {Ronen}}, \
  and\ \bibinfo {author} {\bibfnamefont {J.~L.}\ \bibnamefont {Bohn}},\ }\href
  {\doibase 10.1103/PhysRevA.79.013621} {\bibfield  {journal} {\bibinfo
  {journal} {Phys. Rev. A}\ }\textbf {\bibinfo {volume} {79}},\ \bibinfo
  {pages} {013621} (\bibinfo {year} {2009}{\natexlab{b}})}\BibitemShut
  {NoStop}%
\bibitem [{Note2()}]{Note2}%
  \BibitemOpen
  \bibinfo {note} {For an analogous calculation, but at $T = 0$, see \cite
  {Ronen2006a}}\BibitemShut {NoStop}%
\bibitem [{Note3()}]{Note3}%
  \BibitemOpen
  \bibinfo {note} {For definiteness, this discussion relates to the in-plane
  interaction between atoms in the lowest $z$ vibrational mode.}\BibitemShut
  {Stop}%
\bibitem [{\citenamefont {Fischer}(2006)}]{Fischer2006a}%
  \BibitemOpen
  \bibfield  {author} {\bibinfo {author} {\bibfnamefont {U.~R.}\ \bibnamefont
  {Fischer}},\ }\href {\doibase 10.1103/PhysRevA.73.031602} {\bibfield
  {journal} {\bibinfo  {journal} {Phys. Rev. A}\ }\textbf {\bibinfo {volume}
  {73}},\ \bibinfo {pages} {031602} (\bibinfo {year} {2006})}\BibitemShut
  {NoStop}%
\bibitem [{\citenamefont {M\"uller}\ \emph {et~al.}(2011)\citenamefont
  {M\"uller}, \citenamefont {Billy}, \citenamefont {Henn}, \citenamefont
  {Kadau}, \citenamefont {Griesmaier}, \citenamefont {Jona-Lasinio},
  \citenamefont {Santos},\ and\ \citenamefont {Pfau}}]{Muller2011a}%
  \BibitemOpen
  \bibfield  {author} {\bibinfo {author} {\bibfnamefont {S.}~\bibnamefont
  {M\"uller}}, \bibinfo {author} {\bibfnamefont {J.}~\bibnamefont {Billy}},
  \bibinfo {author} {\bibfnamefont {E.~A.~L.}\ \bibnamefont {Henn}}, \bibinfo
  {author} {\bibfnamefont {H.}~\bibnamefont {Kadau}}, \bibinfo {author}
  {\bibfnamefont {A.}~\bibnamefont {Griesmaier}}, \bibinfo {author}
  {\bibfnamefont {M.}~\bibnamefont {Jona-Lasinio}}, \bibinfo {author}
  {\bibfnamefont {L.}~\bibnamefont {Santos}}, \ and\ \bibinfo {author}
  {\bibfnamefont {T.}~\bibnamefont {Pfau}},\ }\href {\doibase
  10.1103/PhysRevA.84.053601} {\bibfield  {journal} {\bibinfo  {journal} {Phys.
  Rev. A}\ }\textbf {\bibinfo {volume} {84}},\ \bibinfo {pages} {053601}
  (\bibinfo {year} {2011})}\BibitemShut {NoStop}%
\bibitem [{\citenamefont {Santos}\ \emph {et~al.}(2003)\citenamefont {Santos},
  \citenamefont {Shlyapnikov},\ and\ \citenamefont {Lewenstein}}]{Santos2003a}%
  \BibitemOpen
  \bibfield  {author} {\bibinfo {author} {\bibfnamefont {L.}~\bibnamefont
  {Santos}}, \bibinfo {author} {\bibfnamefont {G.~V.}\ \bibnamefont
  {Shlyapnikov}}, \ and\ \bibinfo {author} {\bibfnamefont {M.}~\bibnamefont
  {Lewenstein}},\ }\href {\doibase 10.1103/PhysRevLett.90.250403} {\bibfield
  {journal} {\bibinfo  {journal} {Phys. Rev. Lett.}\ }\textbf {\bibinfo
  {volume} {90}},\ \bibinfo {pages} {250403} (\bibinfo {year}
  {2003})}\BibitemShut {NoStop}%
\bibitem [{\citenamefont {Wilson}\ \emph {et~al.}(2010)\citenamefont {Wilson},
  \citenamefont {Ronen},\ and\ \citenamefont {Bohn}}]{Wilson2010a}%
  \BibitemOpen
  \bibfield  {author} {\bibinfo {author} {\bibfnamefont {R.~M.}\ \bibnamefont
  {Wilson}}, \bibinfo {author} {\bibfnamefont {S.}~\bibnamefont {Ronen}}, \
  and\ \bibinfo {author} {\bibfnamefont {J.~L.}\ \bibnamefont {Bohn}},\ }\href
  {\doibase 10.1103/PhysRevLett.104.094501} {\bibfield  {journal} {\bibinfo
  {journal} {Phys. Rev. Lett.}\ }\textbf {\bibinfo {volume} {104}},\ \bibinfo
  {pages} {094501} (\bibinfo {year} {2010})}\BibitemShut {NoStop}%
\bibitem [{\citenamefont {Ticknor}\ \emph {et~al.}(2011)\citenamefont
  {Ticknor}, \citenamefont {Wilson},\ and\ \citenamefont
  {Bohn}}]{Ticknor2011a}%
  \BibitemOpen
  \bibfield  {author} {\bibinfo {author} {\bibfnamefont {C.}~\bibnamefont
  {Ticknor}}, \bibinfo {author} {\bibfnamefont {R.~M.}\ \bibnamefont {Wilson}},
  \ and\ \bibinfo {author} {\bibfnamefont {J.~L.}\ \bibnamefont {Bohn}},\
  }\href {\doibase 10.1103/PhysRevLett.106.065301} {\bibfield  {journal}
  {\bibinfo  {journal} {Phys. Rev. Lett.}\ }\textbf {\bibinfo {volume} {106}},\
  \bibinfo {pages} {065301} (\bibinfo {year} {2011})}\BibitemShut {NoStop}%
\bibitem [{\citenamefont {Blakie}\ \emph {et~al.}(2012)\citenamefont {Blakie},
  \citenamefont {Baillie},\ and\ \citenamefont {Bisset}}]{Blakie2012a}%
  \BibitemOpen
  \bibfield  {author} {\bibinfo {author} {\bibfnamefont {P.~B.}\ \bibnamefont
  {Blakie}}, \bibinfo {author} {\bibfnamefont {D.}~\bibnamefont {Baillie}}, \
  and\ \bibinfo {author} {\bibfnamefont {R.~N.}\ \bibnamefont {Bisset}},\
  }\href@noop {} {\bibfield  {journal} {\bibinfo  {journal} {arxiv}\ ,\
  \bibinfo {pages} {1206.2770}} (\bibinfo {year} {2012})}\BibitemShut {NoStop}%
\bibitem [{\citenamefont {Dalfovo}\ \emph {et~al.}(1999)\citenamefont
  {Dalfovo}, \citenamefont {Giorgini}, \citenamefont {Pitaevskii},\ and\
  \citenamefont {Stringari}}]{Dalfovo1999}%
  \BibitemOpen
  \bibfield  {author} {\bibinfo {author} {\bibfnamefont {F.}~\bibnamefont
  {Dalfovo}}, \bibinfo {author} {\bibfnamefont {S.}~\bibnamefont {Giorgini}},
  \bibinfo {author} {\bibfnamefont {L.}~\bibnamefont {Pitaevskii}}, \ and\
  \bibinfo {author} {\bibfnamefont {S.}~\bibnamefont {Stringari}},\ }\href@noop
  {} {\bibfield  {journal} {\bibinfo  {journal} {Rev.~Mod.~Phys.}\ }\textbf
  {\bibinfo {volume} {71}},\ \bibinfo {pages} {463} (\bibinfo {year}
  {1999})}\BibitemShut {NoStop}%
\bibitem [{\citenamefont {Glaum}\ \emph {et~al.}(2007)\citenamefont {Glaum},
  \citenamefont {Pelster}, \citenamefont {Kleinert},\ and\ \citenamefont
  {Pfau}}]{Glaum2007}%
  \BibitemOpen
  \bibfield  {author} {\bibinfo {author} {\bibfnamefont {K.}~\bibnamefont
  {Glaum}}, \bibinfo {author} {\bibfnamefont {A.}~\bibnamefont {Pelster}},
  \bibinfo {author} {\bibfnamefont {H.}~\bibnamefont {Kleinert}}, \ and\
  \bibinfo {author} {\bibfnamefont {T.}~\bibnamefont {Pfau}},\ }\href {\doibase
  10.1103/PhysRevLett.98.080407} {\bibfield  {journal} {\bibinfo  {journal}
  {Phys. Rev. Lett.}\ }\textbf {\bibinfo {volume} {98}},\ \bibinfo {pages}
  {080407} (\bibinfo {year} {2007})}\BibitemShut {NoStop}%
\bibitem [{\citenamefont {Glaum}\ and\ \citenamefont
  {Pelster}(2007)}]{Glaum2007A}%
  \BibitemOpen
  \bibfield  {author} {\bibinfo {author} {\bibfnamefont {K.}~\bibnamefont
  {Glaum}}\ and\ \bibinfo {author} {\bibfnamefont {A.}~\bibnamefont
  {Pelster}},\ }\href {\doibase 10.1103/PhysRevA.76.023604} {\bibfield
  {journal} {\bibinfo  {journal} {Phys. Rev. A}\ }\textbf {\bibinfo {volume}
  {76}},\ \bibinfo {pages} {023604} (\bibinfo {year} {2007})}\BibitemShut
  {NoStop}%
\bibitem [{Note4()}]{Note4}%
  \BibitemOpen
  \bibinfo {note} {Note we use fixed geometric mean trap frequency, whereas
  \cite {Ronen2007a} fixes $\omega _{\rho }$. The interaction parameter used in
  \cite {Ronen2007a} ($N\gg 1$ limit) is $D= NC_{\protect \textrm {dd}}/4\pi
  \hbar \omega _\rho a_\rho ^3$) with $a_{\rho }=\protect \sqrt {\hbar /M\omega
  _{\rho }}$, which relates to our parameter as $D=N^{5/6}\lambda
  ^{-1/6}C_{\protect \textrm {dd}}/4\pi C_0 $.}\BibitemShut {Stop}%
\bibitem [{\citenamefont {Blakie}\ \emph {et~al.}(2008)\citenamefont {Blakie},
  \citenamefont {Bradley}, \citenamefont {Davis}, \citenamefont {Ballagh},\
  and\ \citenamefont {Gardiner}}]{cfieldRev2008}%
  \BibitemOpen
  \bibfield  {author} {\bibinfo {author} {\bibfnamefont {P.~B.}\ \bibnamefont
  {Blakie}}, \bibinfo {author} {\bibfnamefont {A.~S.}\ \bibnamefont {Bradley}},
  \bibinfo {author} {\bibfnamefont {M.~J.}\ \bibnamefont {Davis}}, \bibinfo
  {author} {\bibfnamefont {R.~J.}\ \bibnamefont {Ballagh}}, \ and\ \bibinfo
  {author} {\bibfnamefont {C.~W.}\ \bibnamefont {Gardiner}},\ }\href@noop {}
  {\bibfield  {journal} {\bibinfo  {journal} {Adv. Phys.}\ }\textbf {\bibinfo
  {volume} {57}},\ \bibinfo {pages} {363} (\bibinfo {year} {2008})}\BibitemShut
  {NoStop}%
\bibitem [{\citenamefont {Sau}\ \emph {et~al.}(2009)\citenamefont {Sau},
  \citenamefont {Leslie}, \citenamefont {Stamper-Kurn},\ and\ \citenamefont
  {Cohen}}]{Sau2009a}%
  \BibitemOpen
  \bibfield  {author} {\bibinfo {author} {\bibfnamefont {J.~D.}\ \bibnamefont
  {Sau}}, \bibinfo {author} {\bibfnamefont {S.~R.}\ \bibnamefont {Leslie}},
  \bibinfo {author} {\bibfnamefont {D.~M.}\ \bibnamefont {Stamper-Kurn}}, \
  and\ \bibinfo {author} {\bibfnamefont {M.~L.}\ \bibnamefont {Cohen}},\ }\href
  {\doibase 10.1103/PhysRevA.80.023622} {\bibfield  {journal} {\bibinfo
  {journal} {Phys. Rev. A}\ }\textbf {\bibinfo {volume} {80}},\ \bibinfo
  {pages} {023622} (\bibinfo {year} {2009})}\BibitemShut {NoStop}%
\bibitem [{\citenamefont {Blakie}\ \emph {et~al.}(2009)\citenamefont {Blakie},
  \citenamefont {Ticknor}, \citenamefont {Bradley}, \citenamefont {Martin},
  \citenamefont {Davis},\ and\ \citenamefont {Kawaguchi}}]{Blakie2009e}%
  \BibitemOpen
  \bibfield  {author} {\bibinfo {author} {\bibfnamefont {P.~B.}\ \bibnamefont
  {Blakie}}, \bibinfo {author} {\bibfnamefont {C.}~\bibnamefont {Ticknor}},
  \bibinfo {author} {\bibfnamefont {A.~S.}\ \bibnamefont {Bradley}}, \bibinfo
  {author} {\bibfnamefont {A.~M.}\ \bibnamefont {Martin}}, \bibinfo {author}
  {\bibfnamefont {M.~J.}\ \bibnamefont {Davis}}, \ and\ \bibinfo {author}
  {\bibfnamefont {Y.}~\bibnamefont {Kawaguchi}},\ }\href {\doibase
  10.1103/PhysRevE.80.016703} {\bibfield  {journal} {\bibinfo  {journal} {Phys.
  Rev. E}\ }\textbf {\bibinfo {volume} {80}},\ \bibinfo {pages} {016703}
  (\bibinfo {year} {2009})}\BibitemShut {NoStop}%
\bibitem [{\citenamefont {\ifmmode~\acute{S}\else \'{S}\fi{}wis\l{}ocki}\ \emph
  {et~al.}(2010)\citenamefont {\ifmmode~\acute{S}\else \'{S}\fi{}wis\l{}ocki},
  \citenamefont {Brewczyk}, \citenamefont {Gajda},\ and\ \citenamefont
  {Rz\k{a}\ifmmode~\dot{z}\else \.{z}\fi{}ewski}}]{Tomasz2010a}%
  \BibitemOpen
  \bibfield  {author} {\bibinfo {author} {\bibfnamefont {T.}~\bibnamefont
  {\ifmmode~\acute{S}\else \'{S}\fi{}wis\l{}ocki}}, \bibinfo {author}
  {\bibfnamefont {M.}~\bibnamefont {Brewczyk}}, \bibinfo {author}
  {\bibfnamefont {M.}~\bibnamefont {Gajda}}, \ and\ \bibinfo {author}
  {\bibfnamefont {K.}~\bibnamefont {Rz\k{a}\ifmmode~\dot{z}\else
  \.{z}\fi{}ewski}},\ }\href {\doibase 10.1103/PhysRevA.81.033604} {\bibfield
  {journal} {\bibinfo  {journal} {Phys. Rev. A}\ }\textbf {\bibinfo {volume}
  {81}},\ \bibinfo {pages} {033604} (\bibinfo {year} {2010})}\BibitemShut
  {NoStop}%
\bibitem [{\citenamefont {Blaizot}\ and\ \citenamefont
  {Ripka}(1986)}]{BlaizotRipka}%
  \BibitemOpen
  \bibfield  {author} {\bibinfo {author} {\bibfnamefont {J.}~\bibnamefont
  {Blaizot}}\ and\ \bibinfo {author} {\bibfnamefont {G.}~\bibnamefont
  {Ripka}},\ }\href@noop {} {\emph {\bibinfo {title} {{Quantum Theory of Finite
  Systems}}}},\ \bibinfo {edition} {1st}\ ed.\ (\bibinfo  {publisher} {MIT
  Press},\ \bibinfo {address} {Cambridge, Massachusetts},\ \bibinfo {year}
  {1986})\BibitemShut {NoStop}%
\bibitem [{\citenamefont {Castin}\ and\ \citenamefont
  {Dum}(1998)}]{Castin1998a}%
  \BibitemOpen
  \bibfield  {author} {\bibinfo {author} {\bibfnamefont {Y.}~\bibnamefont
  {Castin}}\ and\ \bibinfo {author} {\bibfnamefont {R.}~\bibnamefont {Dum}},\
  }\href {\doibase 10.1103/PhysRevA.57.3008} {\bibfield  {journal} {\bibinfo
  {journal} {Phys. Rev. A}\ }\textbf {\bibinfo {volume} {57}},\ \bibinfo
  {pages} {3008} (\bibinfo {year} {1998})}\BibitemShut {NoStop}%
\bibitem [{\citenamefont {Morgan}(2000)}]{Morgan2000a}%
  \BibitemOpen
  \bibfield  {author} {\bibinfo {author} {\bibfnamefont {S.~A.}\ \bibnamefont
  {Morgan}},\ }\href@noop {} {\bibfield  {journal} {\bibinfo  {journal} {J.
  Phys. B}\ }\textbf {\bibinfo {volume} {33}},\ \bibinfo {pages} {3847}
  (\bibinfo {year} {2000})}\BibitemShut {NoStop}%
\bibitem [{Note5()}]{Note5}%
  \BibitemOpen
  \bibinfo {note} {Note in Bogoliubov theory many practitioners neglect to
  perform the projection, which fortuitously does not change the quasiparticle
  energies. However, in Hartree-Fock theory the mode energies are affected by
  projection.}\BibitemShut {Stop}%
\bibitem [{\citenamefont {Hohenberg}\ and\ \citenamefont
  {Martin}(1965)}]{Hohenberg1965a}%
  \BibitemOpen
  \bibfield  {author} {\bibinfo {author} {\bibfnamefont {P.}~\bibnamefont
  {Hohenberg}}\ and\ \bibinfo {author} {\bibfnamefont {P.}~\bibnamefont
  {Martin}},\ } {\bibfield  {journal}
  {\bibinfo  {journal} {Annals of Physics},\ }\textbf {\bibinfo {volume}
  {34}},\ \bibinfo {pages} {291 } (\bibinfo {year} {1965})},\ ISSN \bibinfo
  {issn} {0003-4916}\BibitemShut {NoStop}%
\bibitem [{\citenamefont {Griffin}(1996)}]{TheoryA6}%
  \BibitemOpen
  \bibfield  {author} {\bibinfo {author} {\bibfnamefont {A.}~\bibnamefont
  {Griffin}},\ }\href@noop {} {\bibfield  {journal} {\bibinfo  {journal} {Phys.
  Rev. B},\ }\textbf {\bibinfo {volume} {53}},\ \bibinfo {pages} {9341}
  (\bibinfo {year} {1996})}\BibitemShut {NoStop}%
\end{thebibliography}

%

\end{document}